\documentclass{article}

\usepackage{PRIMEarxiv}

\usepackage[utf8]{inputenc}
\usepackage[T1]{fontenc}
\usepackage{amsmath,amssymb,amsthm}
\usepackage[numbers,sort&compress]{natbib}
\usepackage[hidelinks,pdftitle={Attention Tax, Handoff Tax: A Stylised Model of When Multi-Agent LLM Systems Help},pdfauthor={Akshit Anchan and Nayonika Sen}]{hyperref}
\usepackage{url}
\usepackage{booktabs}
\usepackage{nicefrac}
\usepackage{microtype}
\usepackage{fancyhdr}
\usepackage{graphicx}
\usepackage{tikz}
\usetikzlibrary{arrows.meta,positioning,calc}
\graphicspath{{media/}}

\theoremstyle{plain}
\newtheorem{theorem}{Theorem}
\newtheorem{corollary}{Corollary}
\newtheorem{proposition}{Proposition}
\theoremstyle{definition}
\newtheorem{assumption}{Assumption}
\newtheorem{definition}{Definition}
\theoremstyle{remark}

\newcommand{\eps}{\varepsilon}
\newcommand{\GSA}{G_{\mathrm{SA}}}
\newcommand{\GD}{G_{\mathrm{D}}}
\newcommand{\Evote}{E_{\mathrm{vote}}}

\title{Attention Tax, Handoff Tax: A Stylised Model of\\ When Multi-Agent LLM Systems Help
\thanks{\textit{\underline{Citation}}:
\textbf{A. P. Anchan and N. Sen. Attention Tax, Handoff Tax: A Stylised Model of When Multi-Agent LLM Systems Help. Preprint, 2026.}}
}

\author{
  Akshit Anchan \\
  University of Amsterdam \\
  Amsterdam, The Netherlands \\
  \texttt{akshit.anchan@student.uva.nl}
  \And
  Nayonika Sen \\
  Northeastern University \\
  Boston, MA, United States of America \\
  \texttt{sen.na@northeastern.edu}
}

\begin{document}
\maketitle

\begin{abstract}
Recent work on multi-agent LLM systems reaches sharply different conclusions: some results show that a single agent with the same information and compute should dominate a delegated system, others that multi-agent gains grow with task depth. We argue that much of the disagreement comes from modelling different bottlenecks, and introduce a stylised reliability model built around two trade-offs. Decomposition reduces the burden of long contexts but incurs a handoff tax when information is compressed or transferred between agents. Redundancy gains from multiple samples, but its benefit depends on how much their failures are shared. With reasoning budget, verification, and task structure added, the model yields two crossover conditions: decomposition becomes preferable once the attention cost avoided by resetting context exceeds the handoff cost, and parallel sampling at equal budget is eventually preferable when its shared-failure floor lies below the error floor of one agent thinking longer. We connect these regimes to recent theoretical and empirical results. On a ledger-reconciliation task we measure the context-degradation curve and the handoff tax from single-agent and handoff runs alone. From these the model places the crossover at depth 10 and predicts decomposition to win at depths 20, 50, and 100. It does, on step-level and final-balance accuracy, and the decomposed system's success, which the prediction never sees, lands within 9 percentage points of the predicted rate at every depth.
\end{abstract}


\section{Introduction}

There is a live argument about whether multi-agent LLM systems are worth building, and it has now been formalised by both sides. \citet{tran2026single} argue from the data processing inequality that under equal reasoning-token budgets a single agent is more information-efficient than any system that passes summaries between agents, and they confirm it on multi-hop question answering across three model families. \citet{ao2026reliability} prove the decision-theoretic version: without new exogenous signals, any delegated network is dominated by a centralised Bayes decision maker with access to the same information. On the other side, \citet{tang2025task} prove that the relative gain of a multi-agent system over a single agent increases in both task depth and task width, and grows without bound in depth. \citet{meyerson2025million} then solve a million-step task with zero errors by decomposing it into single-step subtasks, something no single agent comes close to. And \citet{jwalapuram2026illusion} put both outcomes in one table: a self-consistency baseline outperforms six automated multi-agent frameworks almost everywhere they test, often at under a tenth of the cost, while an expert-designed decomposition raises GPT-5 from 57.0 to 96.5 on a context-heavy task the automated frameworks also fail.

Taken together, the papers cannot all describe the same world, and yet each of them is a theorem. Our claim is that much of the contradiction comes from bookkeeping. Each of these models fixes some of five quantities and lets the others move, and a quantity one side fixes is often exactly the one the other side varies. The sceptics' dominance theorems hold for a single agent whose per-step accuracy does not depend on how much sits in its context, and where they relax that, the relaxation is an exogenous corruption level rather than accumulated load. Positive results likewise use different assumptions about information, context, communication and compute. MAKER explicitly analyses voting costs and passes task state between steps.

This paper writes down a stylised model with all five quantities and works out what happens in it. A task is a chain of $L$ steps. A step is an LLM call whose failure probability $\eps(c,b)$ depends on the context load $c$ it attends over and the thinking budget $b$ it receives. Two structural primitives are available: decomposition, which gives each agent a fresh context at the price of a handoff loss $h$, and redundancy, which runs $N$ parallel calls whose failures are dependent because they share a base model. Verifiers at boundaries recover some errors. Budgets are matched per step.

Each prior result then corresponds to a restriction of the model. We say for each whether the correspondence is exact, structural, or an analogy. The decomposition question has a crossover. For single-step agents there is a unique depth $L^{*}$ past which decomposition wins, and it exists if and only if the curve's asymptotic rise above a fresh agent's operating point exceeds one handoff's cost. For arbitrary chunk sizes the criterion is the total tax a fresh agent avoids over its whole run, which can exceed the handoff cost even when the single-step gap does not. A flat curve gives no crossover at any chunk size, which is the sceptics' conclusion. In the linear regime the crossover and the chunk size have closed forms. The redundancy question compares two floors under a latent-rate mixture model of the samples: parallel sampling at equal budget wins for all sufficiently large sample counts whenever its shared-failure floor, $\rho\,\eps$ in our common-shock example, is below the single agent's saturated thinking error $\eps_\infty$. Equal floors require a finite-budget comparison.

Two measurements accompany the model. The first is a padded-context pilot: a thousand calls to a self-hosted gpt-oss-20b at five context loads, which measures the curve and nothing else, and which showed that the curve's shape depends on what one counts as a step failure. The second is the test the model asks for. On a ledger-reconciliation task with two long-range dependencies, we measured the in-task curve from single-agent runs and the handoff tax from paired continue-versus-handoff runs, and computed from those two measurements alone the depth at which decomposition into five-step agents should start winning and the success rate of each system at every tested depth. The decomposed system won at every depth, as predicted. The predicted crossover sits at depth 10 and the shallowest depth we ran was 20, so the runs test the winner past the crossover and not the crossing itself.

\paragraph{What is proved, what is assumed, what is measured.}
Because the paper mixes four kinds of claim, Table~\ref{tab:guide} sorts them before any symbols appear. Everything in the first row is a theorem inside the model of Section~\ref{sec:model}, and nothing in it is a statement about LLMs except through the assumptions in the second row.

\begin{table}[t]
\caption{Reading guide. Where each kind of claim lives in the paper.}
\label{tab:guide}
\centering
\small
\begin{tabular}{@{}p{3.1cm}p{11.8cm}@{}}
\toprule
Status & Content \\
\midrule
Proved, within the model & Theorems~\ref{thm:crossover} to~\ref{thm:parseq}, Corollaries~\ref{cor:flat} to~\ref{cor:maker}, and Proposition~\ref{prop:summary}. Proposition~\ref{prop:knee} is a derived approximation, numerically checked. Proofs are in Appendix~\ref{app:proofs}, and key closed forms and boundary cases are checked numerically in Appendix~\ref{app:numerics}. \\
Assumed & Assumptions~\ref{as:product} to~\ref{as:handoff} (product form, monotone curve, context growth, handoff monotonicity), a latent-rate mixture model for parallel samples, and the common-shock law used as a worked example. Proposition~\ref{prop:verify} is conditional on an untested hypothesis about verifiers. \\
Measured & A padded-context curve for one model (Section~\ref{sec:calib}). An in-task curve, a handoff tax at three note lengths, and single-agent and decomposed success at three depths on one task with one model (Section~\ref{sec:exp}). \\
Interpreted & The placement of published results in regimes (Section~\ref{sec:contradiction}), the illustrative crossover depths computed from the pilot curve under assumed handoff costs (Section~\ref{sec:calib}), the proposed mechanism for the instruction-compliance decay, and the location of the ledger crossover below depth 20 (Section~\ref{sec:exp}). \\
\bottomrule
\end{tabular}
\end{table}

\section{The three camps, and what each holds fixed}

\paragraph{Sceptics.}
\citet{tran2026single} model the messages between agents as a Markov chain on the task information and apply the data processing inequality. Their dominance result assumes ``perfect context utilization'': the single agent extracts from a long context everything it would extract from a short one. They then relax it. Their Section~3.1 models an effective context degraded by an exogenous corruption level, shows by the same inequality that the single agent's achievable error is monotone in that level, and their Section~5.3 corrupts contexts by masking and substitution and finds the ordering reverse at heavy corruption. What they do not have is degradation as a function of the load a trajectory accumulates, a depth or chunk-size model, or a way to compute where the reversal falls from measured quantities. That is the gap this paper fills, and the sceptic corner of our model is their undegraded case. \citet{ao2026reliability} work at the level of information structure and deliberately not at the level of token generation. Their delegated decision network has nodes with decision rules, a common signal (the shared context), private exogenous signals (tool outputs), and messages along edges. Proposition~6 there says that with no new exogenous signals the network is dominated by a centralised Bayes decision maker, and Theorem~8 says the loss from communication is an expected posterior divergence, equal to the conditional mutual information $I(Y;H\,|\,M)$ under log loss. The centre in their model has no capacity constraint, and a capacity constraint is the only thing our model adds to theirs. The empirical audits lean the same way. \citet{xu2026rethinking} show a single agent with multi-turn interaction matches homogeneous workflows, \citet{orogat2026unified} find that orchestration depth limits scalability independently of the model, \citet{kim2025scaling} fit a predictive model of when adding agents helps (cross-validated $R^{2}=0.37$) and find gains on some task families and losses of up to 70 per cent on sequential planning, and \citet{jwalapuram2026illusion} report that many automated multi-agent frameworks collapse into self-consistency in practice. \citet{cemri2025mast} supply the failure taxonomy behind much of this.

\paragraph{Optimists.}
\citet{tang2025task} define per-step success $\sigma(w)=u^{w}$ for a task of width $w$ (we rename their symbols to keep ours free) and put $S_{\mathrm{single}}(d,w)=\sigma(w)^{d}$ against a multi-agent success in which $N$ agents answer each step independently and an aggregator picks the correct answer with probability $r$ whenever some agent has it. Their printed Definition~2.3 places $r$ outside the $d$-th power, while their Propositions~2.1 and~2.2 and proofs use the per-step form $[\,r(1-(1-\sigma(w))^{N})\,]^{d}$, which we take to be the intended one. Their relative gain increases in $d$ and $w$, saturates at $(rN)^{d}-1$ in width, and is unbounded in depth. Two features matter for what follows. The single agent's per-step success is the same $\sigma(w)$ as each worker's, so their single agent does not degrade with context either. And nothing is budget-normalised: the multi-agent system spends $N$ calls per step at every one of $d$ steps. \citet{meyerson2025million} supply the engineering version of the same idea. They decompose a Towers-of-Hanoi run into a million single-step subtasks with a first-to-ahead-by-$k$ vote at each, cite context growth as the reason a single agent fails, and report a per-step error rate that stays flat as the task scales once each agent sees only what its step needs. \citet{su2026when} study the training-time question and get a PAC dichotomy: independent subtasks make multi-agent fine-tuning more sample-efficient, dependent subtasks add a quadratic penalty from error propagation.

\paragraph{Redundancy and aggregation.}
A third literature never decomposes anything and only asks what $N$ parallel answers buy. \citet{yang2026diversity} bound the information any transcript can carry by $H(Y|X)$ and introduce an effective channel count $K$ with a complementarity rate $\alpha$, under which residual uncertainty contracts like $(1-\alpha)^{K}$. Homogeneous agents stop adding channels, so scaling saturates, and heterogeneous ensembles keep going. \citet{patel2026sixsigma} give the exponential reliability gain of consensus voting under independence. \citet{yang2026auditing} show in an appendix how that gain fails under correlated correctness. \citet{maryanskyy2026selection} derive a crossover in selector quality below which team diversity hurts and above which it helps, which in our notation is a threshold on the selector's recall $r$. \citet{choi2025debate} prove that, under their conditions, debate conserves an agent's expected belief, so debate cannot systematically raise expected belief above what the initial round already carries, and \citet{liu2026martingale} add that correlated errors make debate converge on shared mistakes. None of this is new to statistics. \citet{ladha1992condorcet} worked out the Condorcet jury theorem under correlated votes, and \citet{kuncheva2003diversity} catalogued how ensemble diversity relates to ensemble accuracy. \citet{snell2024scaling} supply the empirical cousin on the single-model side: parallel and sequential test-time compute trade off differently depending on the problem.

\paragraph{Verification and context management.}
\citet{aghazadeh2026reliability} give the most complete reliability theory we know of. A task is a vector of coupled binary subclaims, an architecture is a sparse role-typed factor graph whose check nodes are noisy Boolean verifiers, and three failure modes (an agent abstains, a verifier returns nothing, a message is lost) propagate as set-valued messages under iterative decoding. They prove a density-evolution theorem in the style of \citet{richardson2008modern}, characterise finite-length failures by certificate-stopping sets, and solve cost-constrained architecture optimisation. The theory is erasure-only by design, and their limitations section lists what is outside it: confidently wrong messages, correlated agent failures from shared training data, adaptive routing, and ``compounding prompt-context failures specific to LLM-style verifiers.'' On the context side, \citet{sun2025contextfolding} fold completed sub-trajectories out of the active context and scale a single agent to long horizons that way, which is decomposition without a second agent, and \citet{kang2025acon} and \citet{xia2026contextrot} treat compaction as the lever. Those methods pay our handoff tax. The closest precursor to the decomposition side of our model is \citet{xu2025divide}, who split the error of divide-and-conquer processing of a long input into task noise (cross-chunk dependence), model noise (confusion that grows with context size), and aggregator noise, and show by varying chunk size that chunking pays when model noise grows fast enough and task noise is manageable. Their two leading terms are the parallel-chunking analogues of our handoff tax and curve. \citet{liu2026phase} put finite context, a message-fidelity function, and shared-failure correlation into one budgeted theory and derive phase transitions for star, chain, and tree organisations, with chains degrading unless new evidence enters. Their context limit is a hard window on fan-in and their validation is synthetic. Ours is a graded curve over accumulated load, which is what lets a fresh agent win, and its prediction is tested on an LLM task. What we add is the sequential chain with handoffs, the budget accounting, the dependence and verification axes, the crossover depth with its closed forms, and a prediction checked against a decomposed run. The phrase handoff tax is also used by \citet{ganz2026handoff} for the cost a stronger or cheaper model pays when it takes over another model's trajectory mid-run. Ours is the loss at a summary boundary between agents of the same model, and the two are different quantities.

\paragraph{The curve.}
Whether per-step accuracy depends on context load has a broad but not uniform empirical answer. \citet{liu2024lost} found a U-shaped dependence on where relevant information sits. \citet{hsieh2024ruler} found most long-context models fail well below their advertised length. \citet{modarressi2025nolima} report, in the conference version, eleven of thirteen models falling below half their short-context score at 32k tokens, with GPT-4o dropping from 99.3 to 69.7. \citet{hong2025contextrot} found declining performance with input length across eighteen production models, and also report that the decline is not uniform across tasks and models. \citet{du2025contextlength} isolated the effect from retrieval: accuracy falls with length even when the relevant span is found. \citet{laban2025lost} found the multi-turn version. \citet{sinha2025illusion} found per-step accuracy falling with the number of steps on a long execution task, partly because errors already in context make further errors likelier, an effect an in-task curve such as ours measures together with load. We take this as licence for a monotone first-order model of the curve and nothing stronger, and Section~\ref{sec:calib} shows one case where monotonicity holds for one failure definition and is undetectable for another.

\paragraph{Scaffold effects.}
\citet{zhang2026harness} and \citet{yao2026harnessbench} show that scaffold choice moves agent benchmark scores as much as model choice. Any placement of a published benchmark in our regime map is therefore qualitative, and we do not attempt a quantitative one.

\section{Model}
\label{sec:model}

The model is stylised. It is built to make the five quantities the literature holds fixed into explicit parameters, and the price of that is a first-order treatment of everything else. We state each simplification where it is made. Table~\ref{tab:notation} collects the symbols.

\paragraph{A running example.}
To keep the symbols attached to something, we follow one task through the paper: an agent reconciling a ledger of $L$ transactions, one per step. The task statement is $c_0=1{,}000$ tokens, each transaction adds $k=400$ tokens of tool output, and a summary handed to a fresh agent is $s=300$ tokens. For illustration we take the curve to be linear with $g(c)=0.002+5\times10^{-6}c$, so a fresh agent's per-step failure is about one per cent and a single agent's rises to five per cent by the twenty-second transaction and nineteen per cent by the hundredth. We take the handoff tax to be $h=0.02$ nats with no summary overhead. These are the parameters of Figure~\ref{fig:crossover}. They are chosen to be plausible, not measured, and we say what each theorem means for the ledger as we go.

\begin{table}[t]
\caption{Notation. Loads and lengths are in tokens, budgets in thinking tokens, and losses in nats.}
\label{tab:notation}
\centering
\small
\begin{tabular}{@{}llp{9.2cm}@{}}
\toprule
Symbol & Set in & Meaning \\
\midrule
$L$, $t$ & Sec.~\ref{sec:model} & task depth in steps, and the step index \\
$c_0$, $k$ & Sec.~\ref{sec:model} & task statement size, and tokens added per step \\
$s$, $m$ & Sec.~\ref{sec:model} & summary length, and steps per agent (chunk size) \\
$b$, $o$, $o_v$ & Sec.~\ref{sec:model} & token budget per step, and tokens reserved per summary and per verifier call \\
$\eps(c,b)$, $g(c,b)$ & Eq.~\eqref{eq:g} & per-step failure probability at load $c$ and budget $b$, and its log-loss $-\log(1-\eps)$ \\
$h$, $\Delta_o$, $h_{\mathrm{eff}}$ & Def.~\ref{def:h}, Eqs.~\eqref{eq:deltao}, \eqref{eq:gd} & handoff tax, writer's overhead penalty, and their sum \\
$\GSA$, $\GD$ & Eqs.~\eqref{eq:gsa}, \eqref{eq:gd} & total log-loss of the single agent and of a decomposed system \\
$a_t$, $a_\infty$, $A_\infty$ & Eqs.~\eqref{eq:at}, \eqref{eq:Ainf} & attention tax at step $t$, its limit, and the total tax a fresh agent avoids \\
$D(L)$, $L^{*}$ & Eq.~\eqref{eq:D} & log-loss advantage of decomposition at depth $L$, and the crossover depth \\
$\lambda$, $c_\kappa$, $\lambda_2$ & Sec.~\ref{sec:decomp} & slope of a linear curve, and knee location and post-knee slope of a knee-shaped curve \\
$N$, $\rho$, $\mu$, $F$ & Sec.~\ref{sec:redundancy} & number of parallel samples, their pairwise correlation, the latent-rate mixing law, and the shared-failure mass \\
$q$, $\rho_{gv}$, $q_{\mathrm{eff}}$ & Sec.~\ref{sec:model} & verifier recall, generator-verifier dependence, and effective recall $q(1-\rho_{gv})$ \\
$\eps_b$, $\eps_\infty$ & Sec.~\ref{sec:redundancy} & failure at budget $b$, and the thinking floor as $b\to\infty$ \\
\bottomrule
\end{tabular}
\end{table}

\subsection{Steps, curves, and budgets}

A task is a chain of $L$ steps. Step $t$ consumes the task statement, of size $c_0$, plus everything produced so far. Each step adds $k$ tokens of new material: tool output, observations, retained reasoning. A step is one LLM call. It fails with probability $\eps(c,b)$, where $c$ is the context load it attends over and $b$ is its thinking budget. We work in log-loss units,
\begin{equation}
g(c,b) := -\log\bigl(1-\eps(c,b)\bigr),
\label{eq:g}
\end{equation}
so that products of success probabilities become sums. Here $g$ is the negative log of the probability that the step produces a usable result. It is an operational reliability, not a proper-scoring-rule risk, and we keep it that way throughout.

\begin{assumption}[Product form]
\label{as:product}
Steps fail independently given their $(c,b)$, and the task succeeds only if every step does. Then $-\log S=\sum_t g(c_t,b_t)$.
\end{assumption}

This is the first-order choice that \citet{aghazadeh2026reliability} also make when they take failures conditionally independent given the graph. Within-trajectory dependence is real and is discussed in Section~\ref{sec:limits}. Dependence between parallel agents is treated explicitly in Section~\ref{sec:redundancy}.

\begin{assumption}[Monotone curve]
\label{as:curve}
$\eps$ is nondecreasing in $c$ and nonincreasing in $b$, with $\eps<1$ on the loads and budgets considered, so that $g$ is finite there.
\end{assumption}

We call $c\mapsto\eps(c,b)$ the \emph{context degradation curve}. A flat curve is the case in which our model reaches the dominance conclusions of \citet{tran2026single} and \citet{ao2026reliability}. The linear and knee-shaped forms used later are local descriptions on the range of loads a trajectory actually visits. The curves measured in Sections~\ref{sec:calib} and~\ref{sec:exp} are neither linear nor knee-shaped over their whole range, and the closed forms are only as good as the local fit.

\begin{assumption}[Context growth]
\label{as:growth}
A single agent carries its whole trajectory, so $c_t=c_0+kt$ with $k>0$. A decomposed system with chunk size $m\in\{1,\dots,L\}$ gives each agent up to $m$ consecutive steps in a fresh context. The first agent starts from the task statement alone. Every later agent starts from a summary of length $s$ written by its predecessor, so on its $i$-th step it sees $c_0+s+ki$. Write $L=nm+\bar m$ with $0\le\bar m<m$: there are $n$ full chunks, one final chunk of $\bar m$ steps when $\bar m>0$, and $\lceil L/m\rceil-1$ handoffs.
\end{assumption}

\paragraph{Budget.}
Both systems receive the same thinking budget $b$ per step, so equal total budgets of $bL$. The decomposed system also spends $o$ tokens writing each summary, and we charge that cost to the agent that writes it: an agent that hands off works with $b-o/m$ per step over its $m$ steps, which raises its chunk's log-loss by
\begin{equation}
\Delta_o:=\sum_{i=1}^{m}\bigl[g(c_0+s+ki,\,b-o/m)-g(c_0+s+ki,\,b)\bigr]\ge 0 .
\label{eq:deltao}
\end{equation}
An agent that does not hand off, including the only agent when $m=L$, pays nothing. This puts the overhead where it is incurred, keeps every step's budget independent of $L$, and makes the decomposed system coincide with the single agent whenever there are no handoffs. We evaluate the first writer's penalty at the same loads as the others rather than at its own slightly smaller ones, a second-order simplification that lets $\Delta_o$ be one number per handoff. When verifiers are added, Corollary~\ref{cor:verified} reserves their tokens, and those of a retry, explicitly within the same total.

\subsection{Handoff loss}

When an agent starts a chunk from a summary $M$ instead of the full upstream workspace $H$, its steps can fail more often. We define the handoff tax operationally in the same units as everything else.

\begin{definition}[Handoff tax]
\label{def:h}
Let $G_M$ be the log-loss of a chunk started from the summary $M$, at loads $c_0+s+ki$, and $G_H$ the log-loss of the same chunk started from the full upstream workspace $H$, at loads $c_0+|H|+ki$. The handoff tax is
\begin{equation}
h:=G_M-G_H+\sum_{i=1}^{m}\bigl[g(c_0+|H|+ki,\,b)-g(c_0+s+ki,\,b)\bigr],
\label{eq:hdef}
\end{equation}
the difference the summary makes once the attention tax that the full workspace would have paid, given by the curve, is added back. It is the part of the summary's effect that is not explained by load. Nothing in \eqref{eq:hdef} forces its sign, which is why Assumption~\ref{as:handoff} imposes one.
\end{definition}

Everything on the right of \eqref{eq:hdef} is measurable: run the chunk from $M$ and from $H$ and measure both failure rates, and take the curve term from an independently measured context degradation curve. Under a flat curve $h=G_M-G_H$ is simply the log-loss gap between the two starts. The definition is counterfactual only in the sense that it separates two effects that occur together.

\begin{assumption}[Handoff]
\label{as:handoff}
$h\ge 0$: once load is accounted for, a summary does not improve on the full workspace. And $h$ is nonincreasing in the summary length $s$ whenever longer summaries can carry everything shorter ones do (nested or optimised summarisers). For fixed $s$ and $m$ the same $h$ is paid at every boundary, including one that precedes a shorter final chunk.
\end{assumption}

The relation to \citet{ao2026reliability} is by analogy, not identity. Their communication loss is a gap in Bayes risk under a proper scoring rule, equal to $I(Y;H\,|\,M)$ under log score, for a Bayes-optimal decision maker. Ours is a gap in the negative log of an operational success probability for whatever agent is actually running. The two are both nonnegative, both vanish when the summary is sufficient, and both shrink as the summary grows under the same nesting condition, which is what the data processing inequality gives in their setting. They are not numerically equal, and we do not add one to the other.

\subsection{The two log-losses}

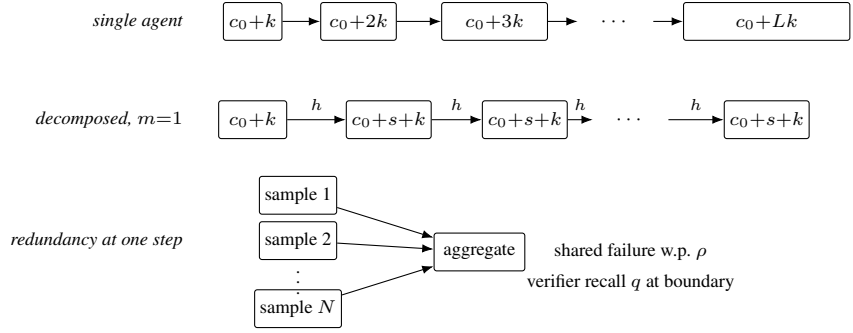
\begin{figure}[t]
\centering
\begin{tikzpicture}[
  box/.style={draw, rounded corners=1pt, minimum height=5mm, inner sep=2pt, font=\scriptsize},
  lab/.style={font=\scriptsize\itshape},
  arr/.style={-{Latex[length=1.5mm]}, thin}
]
\node[lab] at (-0.2,0) [anchor=east] {single agent};
\node[box, minimum width=6mm]  (a1) at (0.6,0) {$c_0{+}k$};
\node[box, minimum width=10mm] (a2) at (2.0,0) {$c_0{+}2k$};
\node[box, minimum width=14mm] (a3) at (3.8,0) {$c_0{+}3k$};
\node[font=\scriptsize] at (5.4,0) {$\cdots$};
\node[box, minimum width=22mm] (aL) at (7.4,0) {$c_0{+}Lk$};
\draw[arr] (a1) -- (a2); \draw[arr] (a2) -- (a3); \draw[arr] (a3) -- (4.9,0); \draw[arr] (5.9,0) -- (aL);
\node[lab] at (-0.2,-1.3) [anchor=east] {decomposed, $m{=}1$};
\node[box, minimum width=9mm] (d1) at (0.6,-1.3) {$c_0{+}k$};
\node[box, minimum width=9mm] (d2) at (2.4,-1.3) {$c_0{+}s{+}k$};
\node[box, minimum width=9mm] (d3) at (4.2,-1.3) {$c_0{+}s{+}k$};
\node[font=\scriptsize] at (5.6,-1.3) {$\cdots$};
\node[box, minimum width=9mm] (dL) at (7.4,-1.3) {$c_0{+}s{+}k$};
\draw[arr] (d1) -- node[above,font=\tiny] {$h$} (d2);
\draw[arr] (d2) -- node[above,font=\tiny] {$h$} (d3);
\draw[arr] (d3) -- node[above,font=\tiny] {$h$} (5.1,-1.3);
\draw[arr] (6.1,-1.3) -- node[above,font=\tiny] {$h$} (dL);
\node[lab] at (-0.2,-2.9) [anchor=east] {redundancy at one step};
\node[box, minimum width=9mm] (r1) at (1.2,-2.3) {sample 1};
\node[box, minimum width=9mm] (r2) at (1.2,-2.9) {sample 2};
\node[font=\scriptsize] at (1.2,-3.35) {$\vdots$};
\node[box, minimum width=9mm] (rN) at (1.2,-3.8) {sample $N$};
\node[box, minimum width=12mm] (agg) at (3.6,-3.05) {aggregate};
\draw[arr] (r1) -- (agg); \draw[arr] (r2) -- (agg); \draw[arr] (rN) -- (agg);
\node[font=\scriptsize] at (5.6,-3.05) {shared failure w.p.\ $\rho$};
\node[font=\scriptsize] at (5.6,-3.45) {verifier recall $q$ at boundary};
\end{tikzpicture}
\caption{The three architectures the model expresses. The single agent's context grows by $k$ per step and its per-step failure follows the curve $\eps(c,b)$. A decomposed system starts its first agent from the task statement, resets context at each boundary, and pays a handoff tax $h$ there. A redundancy node runs $N$ samples whose failures are dependent and combines them with a vote or a selector.}
\label{fig:schematic}
\end{figure}

Write $U_j=\sum_{i=1}^{j}g(c_0+ki,b)$ for a chunk that starts from the task statement and $V_j=\sum_{i=1}^{j}g(c_0+s+ki,b)$ for one that starts from a summary, with $U_0=V_0=0$. Assumptions~\ref{as:product} to~\ref{as:handoff} give
\begin{align}
\GSA(L) &= \sum_{t=1}^{L} g(c_0+kt,\,b), \label{eq:gsa}\\
\GD(L,m) &= U_m+(n-1)\bigl(V_m+h_{\mathrm{eff}}\bigr)+\mathbf{1}_{\{\bar m>0\}}\bigl(V_{\bar m}+h_{\mathrm{eff}}\bigr),\qquad h_{\mathrm{eff}}:=h+\Delta_o , \label{eq:gd}
\end{align}
which is $U_m+(L/m-1)(V_m+h_{\mathrm{eff}})$ when $m$ divides $L$. The first term is the first agent, which has no summary. Every handoff costs $h_{\mathrm{eff}}$, the information loss plus the writer's overhead. At $m=L$ there are no handoffs, the decomposed system is one agent doing everything, and \eqref{eq:gd} reduces to \eqref{eq:gsa} for every $o$, as it should. Decomposition beats the single agent exactly when $\GD<\GSA$. Figure~\ref{fig:schematic} draws the three architectures the model expresses.

\subsection{Redundancy and verification}

At any step the system may instead run $N$ parallel calls and aggregate. Parallel calls to the same base model do not fail independently. We assume the $N$ correctness indicators are generated by an exchangeable Bernoulli mixture: conditionally i.i.d.\ given a latent per-item failure rate $p$ drawn from some law $\mu$ (Section~\ref{sec:redundancy}). For infinite exchangeable sequences this is de Finetti's representation, and for the finite samples we actually draw it is an assumption. The worked example throughout is the common-shock law: with probability $\rho$ a single latent draw $Z\sim\mathrm{Bernoulli}(1-\eps)$ is copied to all $N$ outputs, and otherwise each output is an independent $\mathrm{Bernoulli}(1-\eps)$. A one-line computation gives $\mathrm{Cov}(X_i,X_j)=\rho\,\eps(1-\eps)$, so the pairwise correlation is exactly $\rho$. Aggregation is either a majority vote on correctness or, following \citet{tang2025task}, a selector that returns the correct answer with probability $r$ whenever some sample has it.

A verifier at a boundary checks the fresh agent's step before it is passed on, with recall $q$ on true failures, and never turns a correct output into an incorrect one. Generator and verifier can share the failure's cause, and we let $\rho_{gv}$ be the probability that they do, so effective recall is $q_{\mathrm{eff}}=q(1-\rho_{gv})$. A detected failure triggers one retry, which fails with probability $e_r$ given the original failure and its detection. A step that fails with probability $e$ then fails with probability
\begin{equation}
T(e,q_{\mathrm{eff}},e_r)=e\,(1-q_{\mathrm{eff}})+e\,q_{\mathrm{eff}}\,e_r ,
\label{eq:retry}
\end{equation}
and $e_r=e$ gives the familiar $e(1-q_{\mathrm{eff}})+q_{\mathrm{eff}}e^{2}$ for a retry as good as the first attempt.

\section{The decomposition crossover}
\label{sec:decomp}

Fix $m=1$, with $b$, $s$, and $h_{\mathrm{eff}}$ fixed as above, and define the \emph{attention tax} at step $t$,
\begin{equation}
a_t := g(c_0+kt,\,b)-g(c_0+s+k,\,b),
\label{eq:at}
\end{equation}
the extra log-loss the single agent pays at step $t$ relative to a fresh agent that started from a summary. By Assumption~\ref{as:curve} the sequence $a_t$ is nondecreasing in $t$. Let $a_\infty:=\lim_{t\to\infty}a_t\in(-\infty,+\infty]$, which exists by monotonicity. From \eqref{eq:gsa} and \eqref{eq:gd},
\begin{equation}
D(L):=\GSA(L)-\GD(L,1)=\sum_{t=2}^{L}a_t-(L-1)\,h_{\mathrm{eff}},\qquad D(1)=0 .
\label{eq:D}
\end{equation}
Decomposition wins at depth $L$ iff $D(L)>0$. When $o=0$, $h_{\mathrm{eff}}=h$, and we write $h$ in that case.

Single-step decomposition eventually wins exactly when the single agent's extra per-step cost, far into the run, exceeds one handoff.

\begin{theorem}[Crossover for single-step agents]
\label{thm:crossover}
Under Assumptions~\ref{as:product} to~\ref{as:handoff}, with $b$, $s$, $h_{\mathrm{eff}}$ fixed and $m=1$:
\begin{enumerate}
\item[(i)] If $a_\infty>h_{\mathrm{eff}}$, there is a unique $L^{*}\ge 2$ such that $D(L)>0$ if and only if $L\ge L^{*}$.
\item[(ii)] If $a_\infty\le h_{\mathrm{eff}}$, then $D(L)\le 0$ for every $L$: single-step decomposition never beats the single agent.
\item[(iii)] $L^{*}$ is nondecreasing in $h_{\mathrm{eff}}$, nondecreasing in $s$ at fixed $h_{\mathrm{eff}}$, and nonincreasing under any change to the curve that leaves it unchanged on $[0,c_0+s+k]$ and raises it above that point.
\end{enumerate}
\end{theorem}

The proof is in Appendix~\ref{app:proofs}. The mechanism is that $D$ has nondecreasing increments $a_{L+1}-h_{\mathrm{eff}}$, so it is discretely convex, and $D(1)=0$. A convex sequence that starts at zero crosses zero from below at most once, and does so if and only if its increments eventually turn positive. Part (iii) is stated for changes above the fresh agent's operating point because a change that also raises the curve below that point raises the fresh agent's own log-loss and can move $L^{*}$ either way.

\begin{corollary}[The sceptics' conclusion is the flat corner]
\label{cor:flat}
If $\eps$ does not depend on $c$, then $\GD(L,m)\ge\GSA(L)$ for every $L$, every $m$, and every $h\ge 0$, $o\ge 0$: the single agent weakly dominates every decomposition at every depth.
\end{corollary}

For the ledger this says that if the agent's accuracy did not depend on how many transactions it was carrying, one agent should reconcile ledgers of any length and every handoff would be a pure loss. This is the conclusion of \citet{tran2026single} under their stated premise and of \citet{ao2026reliability} for a capacity-free centre, reached here by bookkeeping rather than by an information inequality. The proof is one line: under a flat curve every decomposed step costs at least what the corresponding single-agent step costs, and the handoffs cost extra. We also checked it exhaustively on a grid.

Theorem~\ref{thm:crossover}(ii) is a statement about single-step agents only, and it does not extend to larger chunks. A saturating curve can make every single handoff a losing trade while a longer chunk, which pays the same handoff once and avoids the attention tax on many steps, still wins. The general criterion is the total tax a fresh agent avoids.

\begin{theorem}[Crossover for arbitrary chunk sizes]
\label{thm:chunks}
Let the curve saturate, $g(c,b)\to g_{\max}(b)<\infty$ as $c\to\infty$, with $o=0$ and the same handoff tax $h$ at every chunk size, and define the \emph{total avoidable tax}
\begin{equation}
A_\infty:=\sum_{i=1}^{\infty}\bigl[g_{\max}(b)-g(c_0+s+ki,\,b)\bigr]\in[0,\infty].
\label{eq:Ainf}
\end{equation}
Then some chunk size $m$ beats the single agent for all sufficiently large $L$ if and only if $A_\infty>h$. Chunk size $m$ in particular does so if and only if $\sum_{i=1}^{m}[g_{\max}-g(c_0+s+ki)]>h$. If $g$ is unbounded, decomposition at every chunk size eventually wins for every finite $h$.
\end{theorem}

For $m=1$ the partial sum is $a_\infty$ and the theorem reduces to Theorem~\ref{thm:crossover}. On a saturating exponential curve with $a_\infty<h<A_\infty$, single-step decomposition never wins while $m=2$ first wins at depth 26 and wins at every depth from 28 (Appendix~\ref{app:numerics}). The relevant object is the area between the saturation level and the curve above the fresh agent's operating point, not the gap at the first step. A model whose accuracy degrades by only a little with context can still justify decomposition if that little accumulates over enough steps.

\subsection{Closed forms in the linear regime}

Suppose $g(c,b)=g_0+\lambda c$ with $\lambda>0$ on the loads both systems visit, and $o=0$ so $h_{\mathrm{eff}}=h$. Then $a_t=\lambda\bigl(k(t-1)-s\bigr)$ and
\begin{equation}
D(L)=(L-1)\Bigl(\tfrac{1}{2}\lambda kL-\lambda s-h\Bigr).
\label{eq:Dlinear}
\end{equation}

\begin{theorem}[Crossover depth, linear regime]
\label{thm:lstar}
Under the conditions above and for $L\ge 2$, $D(L)>0$ if and only if $L>2(h+\lambda s)/(\lambda k)$, so
\begin{equation}
L^{*}=\max\Bigl\{2,\;\Bigl\lfloor\frac{2(h+\lambda s)}{\lambda k}\Bigr\rfloor+1\Bigr\}.
\label{eq:lstar}
\end{equation}
\end{theorem}

\begin{figure}[t]
\centering
\includegraphics[width=0.98\linewidth]{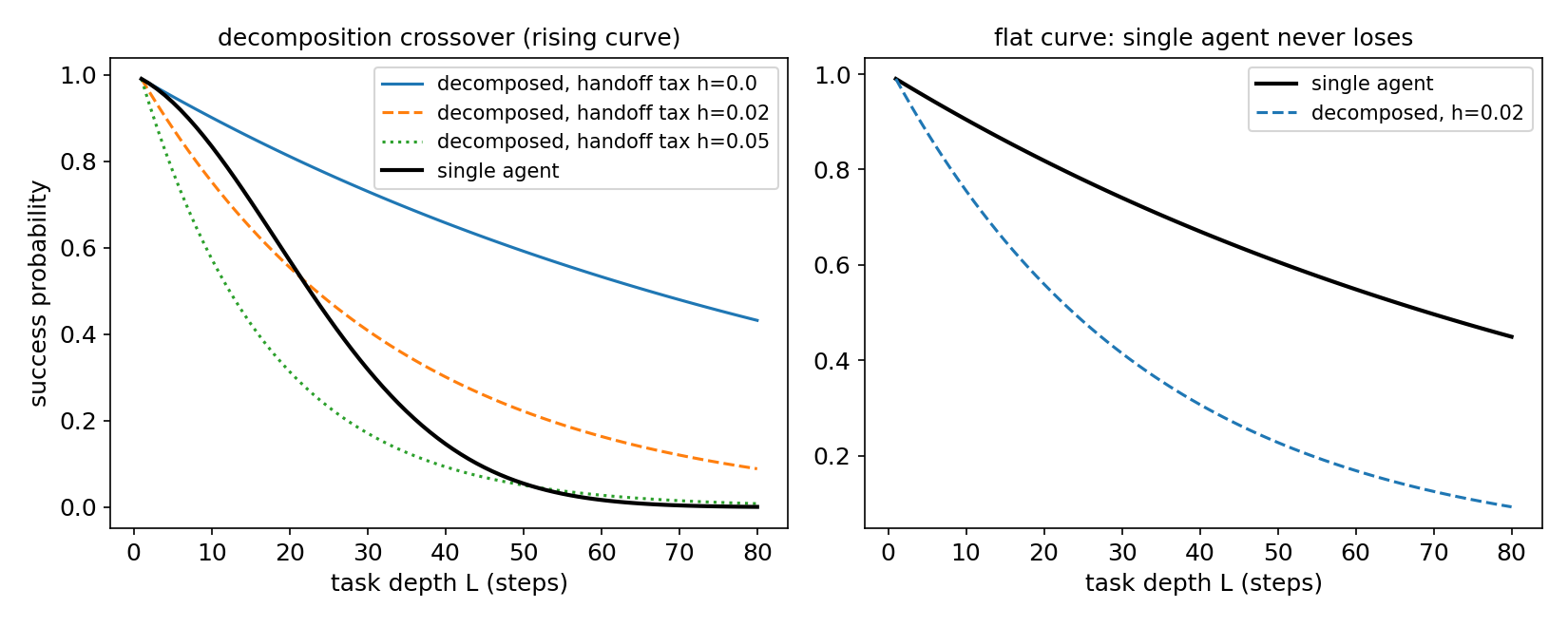}
\caption{Success probability against depth for a single agent (black) and a single-step-per-agent decomposition at three handoff taxes. Left: a linear curve with $g(c)=0.002+5\times10^{-6}c$, $k=400$, $s=300$, $c_0=1000$, $o=0$. Decomposition first wins at $L=22$ for $h=0.02$ and $L=52$ for $h=0.05$, as \eqref{eq:lstar} gives. Right: a flat curve, where the single agent never loses (Corollary~\ref{cor:flat}). Parameters are illustrative, not fitted.}
\label{fig:crossover}
\end{figure}

This is exact. Brute-force evaluation of $D$ against \eqref{eq:lstar} over 200 random parameter draws agrees on every draw. For the ledger, $2(0.02+0.0015)/0.002=21.5$, so $L^{*}=22$: one agent should reconcile ledgers of up to 21 transactions, and past that a chain of fresh agents handing off after every transaction does better, as the left panel of Figure~\ref{fig:crossover} shows. A steeper curve or faster context growth brings the crossover earlier, a costlier handoff pushes it later, and longer summaries push it later through the fresh agents' own load.

\begin{proposition}[Knee-shaped curves]
\label{prop:knee}
Suppose $g(c)=g_0+\lambda_2\max(0,c-c_\kappa)$ on the loads visited, with $\lambda_2>0$, $h>0$, the fresh agents operating below the knee, $c_0+s+k\le c_\kappa$, and $o=0$. Let $T_\kappa=(c_\kappa-c_0)/k$ be the step at which the single agent's context reaches the knee. Replacing the sum of post-knee taxes by an integral gives
\begin{equation}
L^{*}\approx T_\kappa+\frac{h+\sqrt{h^{2}+2\lambda_2 k\,h\,T_\kappa}}{\lambda_2 k}.
\label{eq:knee}
\end{equation}
\end{proposition}

The single agent pays nothing until it crosses the knee, then pays a growing tax, and the crossover sits some way past the knee because the handoffs before it were pure cost. Against brute-force evaluation of \eqref{eq:D} with the parameters fitted in Section~\ref{sec:calib}, \eqref{eq:knee} is within one step in all four cases reported there. We include this case because it is the shape the pilot suggests, with the caveats given there.

\begin{theorem}[Chunk size, linear regime]
\label{thm:chunk}
Under the linear regime, for $m$ dividing $L$,
\begin{equation}
\GD(L,m)=L\bigl[g_0+\lambda c_0\bigr]+\lambda s\,(L-m)+\tfrac{1}{2}L\lambda k(m+1)+\Bigl(\frac{L}{m}-1\Bigr)h .
\label{eq:gdm}
\end{equation}
Let $\alpha=\lambda(k-2s/L)$. For $h>0$ and $\alpha>0$ the continuous relaxation of the per-step cost is strictly convex with minimiser
\begin{equation}
m^{*}=\sqrt{\frac{2h}{\lambda\bigl(k-2s/L\bigr)}}\;\xrightarrow[L\to\infty]{}\;\sqrt{\frac{2h}{\lambda k}},
\label{eq:mstar}
\end{equation}
clipped to $[1,L]$. For $h>0$ and $\alpha\le 0$ the minimiser is $L$, and for $h=0$ the relaxation is linear in $m$ and is minimised at an endpoint. Among divisors of $L$ the optimum is one of the two divisors that bracket $m^{*}$. For a general integer $m$ the final chunk is shorter, \eqref{eq:gd} is evaluated directly, and the integer optimum need not neighbour $m^{*}$. The result treats the per-boundary handoff tax as fixed in $m$. If $h=h(m)$, because a summary must support more downstream work when chunks are longer, the measured objective is minimised directly. With summary overhead $o>0$ and a budget sensitivity $|\partial g/\partial b|\approx\beta$ that is roughly constant over the chunk's loads, $\Delta_o\approx\beta o$ to first order and the same formula holds with $h$ replaced by $h+\beta o$.
\end{theorem}

A grid minimiser of the continuous objective agrees with the $L\to\infty$ form to a relative error of $1.7\times10^{-5}$ on the 91 of 100 random draws where $m^{*}\ge 1$. For the ledger, $\sqrt{0.04/0.002}=4.47$, so each agent should own a few transactions rather than one: handing off after every transaction, as in Figure~\ref{fig:crossover}, pays the handoff tax more often than it needs to. At short depths the rounding matters. With $L=4$, $g(c)=0.01c$, $c_0=s=0$, $k=1$, and $h=0.044$, the losses for $m=1,2,3,4$ are $0.172$, $0.104$, $0.114$, and $0.100$, so the optimum is $m=4$ although $m^{*}=2.97$. The square-root law governs long tasks, and \eqref{eq:gd} settles short ones. The number of steps a subagent should own is the square root of the handoff-to-attention ratio, with overhead entering as an addition to the handoff tax. Cheap handoffs and steep curves push towards one step per agent. Expensive handoffs or flat curves push towards one agent for everything. The folklore that hierarchies should be flattened corresponds to $h_{\mathrm{eff}}$ being large relative to $\lambda k$.

\citet{meyerson2025million} derive a chunk-size law on the cost side that belongs next to this one. Under per-chunk voting, the samples of an $m$-step chunk must agree exactly, so their expected cost grows like $p^{-m}$ in the per-step success $p$, and log-linearly in the number of steps at $m=1$. Equation~\eqref{eq:mstar} is the accuracy-side optimum without per-chunk voting. When both mechanisms are present the joint optimum minimises accuracy loss plus cost over $m$, and the two laws pull the same way.

\begin{proposition}[Summary length]
\label{prop:summary}
Let $h(s)$ be differentiable and strictly convex in the summary length, nonincreasing by Assumption~\ref{as:handoff}, and suppose an interior minimiser exists. For $m=1$ the fresh agent's per-step cost is $g(c_0+s+k)+h(s)$, and in the linear regime its unique minimiser satisfies $-h'(s^{*})=\lambda$.
\end{proposition}

Lengthen the summary until the information it saves per token equals the attention tax per token. Strict convexity is what makes the optimum unique. Monotonicity alone does not, and a summariser with threshold behaviour (nothing useful until a key fact fits) can have several local optima.

\begin{corollary}[Verified boundaries]
\label{cor:verified}
Take $m=1$. The first step receives budget $b$ and is unverified. Each later boundary divides its budget $b$ among the summary, the fresh agent's generation, the verifier, and one retry,
\[
o+b_g+o_v+b_r=b,\qquad o,b_g,o_v,b_r\ge 0,
\]
so the total is $bL$ even when every retry is used. Put $c_f=c_0+s+k$, $e=\eps(c_f,b_g)$, let $e_r$ be the retry's conditional failure at budget $b_r$, and let $\ell_v=-\log\bigl(1-T(e,q_{\mathrm{eff}},e_r)\bigr)$ with $T$ from \eqref{eq:retry}. If verification leaves the handoff tax $h$ unchanged, the verified system has log-loss
\[
G_{\mathrm{D},v}(L)=g(c_0+k,b)+(L-1)(\ell_v+h),
\]
and Theorem~\ref{thm:crossover} holds with $a_t$ replaced by $a_t^{(v)}=g(c_0+kt,b)-\ell_v$ and $h_{\mathrm{eff}}$ by $h$. Against the unverified system with the same summary reservation, whose fresh agents generate at budget $b-o$, verification lowers log-loss at every depth and cannot delay the crossover whenever $\ell_v\le g(c_f,b-o)$.
\end{corollary}

This is how verification enters the crossover: as a reduction of the fresh agent's per-step failure, paid for out of the same budget. Recall alone does not settle it, because the condition charges the verifier and the retry against the tokens the fresh agent would otherwise have spent generating. Section~\ref{sec:verify} takes up how large $q_{\mathrm{eff}}$ is at a boundary.

\subsection{Width is context management}

Give the task $W$ independent branches of depth $L$ whose results are combined at the end. If one agent processes all of them in a single growing context, its load reaches $c_0+kWL$ and, in the linear regime, its excess log-loss grows like $\tfrac{1}{2}\lambda k(WL)^{2}$ while a system that runs each branch in a fresh context pays linearly in $W$. So the multi-agent advantage grows with width, as \citet{tang2025task} find, and depth dominates because depth compounds inside a branch while width only multiplies branches. But the comparison is not really about the number of agents. A single agent that resets its context between branches, as \citet{sun2025contextfolding} implement, is in this model a decomposed system with $W$ chunks and $h$ paid once per branch. The width result is a statement about context management, and the second agent is incidental to it. Tang et al.'s own width effect works through a different channel, a lower per-step success $u^{w}$ that raises the value of redundancy, and we return to it in Section~\ref{sec:reductions}.

\section{The redundancy crossover}
\label{sec:redundancy}

Now hold $L=1$ and ask what $N$ parallel samples buy against one agent that thinks $N$ times as long.

\begin{theorem}[Majority vote under exchangeable failures]
\label{thm:vote}
Let the correctness indicators of $N$ samples be conditionally i.i.d.\ given a latent failure rate $p\sim\mu$, and let $\bar\eps=\int p\,d\mu<\nicefrac{1}{2}$ be the marginal failure rate. For odd $N$ the majority-vote failure probability is
\begin{equation}
\Evote(N)=\int \Pr\bigl[\mathrm{Bin}(N,1-p)\le\tfrac{N-1}{2}\bigr]\,d\mu(p),
\label{eq:vote}
\end{equation}
which converges as $N\to\infty$ to $\mu(p>\nicefrac{1}{2})+\tfrac{1}{2}\mu(p=\nicefrac{1}{2})$, the mass $\mu$ places on shared failure. Under the common-shock law this is
\begin{equation}
\Evote(N)=\rho\,\eps+(1-\rho)\Pr\bigl[\mathrm{Bin}(N,1-\eps)\le\tfrac{N-1}{2}\bigr]\;\xrightarrow[N\to\infty]{}\;\rho\,\eps,
\label{eq:voteshock}
\end{equation}
nonincreasing in $N$ for $\eps<\nicefrac{1}{2}$ and exponentially decaying when $\rho=0$. Under the any-correct selector with recall $r$ the common-shock failure is $1-r\bigl[\rho(1-\eps)+(1-\rho)(1-\eps^{N})\bigr]$, with limit $(1-r)+r\rho\,\eps$.
\end{theorem}

The large-$N$ floor is a property of the whole law $\mu$, not of the pairwise correlation: two laws with the same $\rho$ can have different masses above one half, so ``$\rho\,\eps$'' is the common-shock floor and not a general one. And the vote is modelled on the correctness indicator, which for an open-ended answer with a unique correct value is the pessimistic case: if wrong answers split among several values, plurality over answers succeeds whenever the majority on the indicator does and sometimes when it does not. With those cautions, \eqref{eq:voteshock} contains the exponential gain of \citet{patel2026sixsigma} at $\rho=0$, the failure of voting under correlated correctness in \citet{yang2026auditing}, and the consensus-on-mistakes of \citet{liu2026martingale}. It reproduces the saturation of homogeneous scaling in \citet{yang2026diversity} at the level of mechanism, in that additional samples stop helping once the shared-failure mass dominates, but their effective-channel count is not a function of $\rho$ and we do not claim their theorem as a special case. \citet{tang2025task} sit at $\rho=0$ with the selector, where the per-step failure tends to $1-r$ as $N$ grows.

\begin{corollary}[What pairwise correlation does determine]
\label{cor:cantelli}
For indicators that are conditionally i.i.d.\ given a latent rate, with marginal failure $\eps<\nicefrac12$ and pairwise correlation $\rho$, the latent rate has variance $\sigma^{2}=\rho\,\eps(1-\eps)$, and the shared-failure mass obeys
\begin{equation}
0\;\le\;F\;\le\;F_{\max}(\eps,\rho):=\frac{\rho\,\eps(1-\eps)}{\rho\,\eps(1-\eps)+(\nicefrac12-\eps)^{2}} ,
\label{eq:cantelli}
\end{equation}
with the upper bound approached by two-point laws that put an atom just above $\nicefrac12$ whenever $\rho\le(\nicefrac12-\eps)/(1-\eps)$, and valid but not tight for larger $\rho$. The common-shock floor $\rho\,\eps$ lies within this range.
\end{corollary}

So a measured pair $(\eps,\rho)$ does not pin the floor, but it does bound it, and the bound is what a practitioner can use: if $F_{\max}(\eps_b,\rho)<\eps_\infty$, parallel sampling wins for large $N$ under every mixture law with those moments, and no assumption about the shape of dependence is needed. For one ledger transaction with a 5.5 per cent failure rate and a pairwise correlation of $0.12$ between samples (numbers of the order the pilot saw, used here only for illustration), the common-shock floor is $0.7$ per cent and the bound is $3.1$ per cent. Parallel sampling then beats longer thinking for large $N$ whenever the model's thinking floor on that step is above $3.1$ per cent, whatever the dependence looks like, and above $0.7$ per cent if the dependence is a shared shock.

Parallel sampling beats longer thinking at equal budget for all sufficiently large sample counts whenever the samples' shared-failure floor is below one agent's thinking floor.

\begin{theorem}[Parallel against sequential at equal budget]
\label{thm:parseq}
Let the thinking curve saturate, $\eps(c,b)\to\eps_\infty(c)$ as $b\to\infty$, and write $\eps_b:=\eps(c,b)<\nicefrac{1}{2}$ for the marginal failure at budget $b$. Compare $N$ samples at budget $b$ each with one call at budget $Nb$. As $N\to\infty$ the vote failure tends to the shared-failure mass $F:=\mu(p>\nicefrac{1}{2})+\tfrac12\mu(p=\nicefrac12)$ and the sequential failure tends to $\eps_\infty(c)$. Hence parallel wins for all large $N$ if $F<\eps_\infty(c)$, and sequential wins for all large $N$ if $F>\eps_\infty(c)$. A sufficient condition for parallel to win that uses only $(\eps_b,\rho)$ is $F_{\max}(\eps_b,\rho)<\eps_\infty(c)$. Under the common-shock law $F=\rho\,\eps_b$ and the threshold correlation is $\rho^{*}=\eps_\infty(c)/\eps_b$.
\end{theorem}

\begin{figure}[t]
\centering
\includegraphics[width=0.62\linewidth]{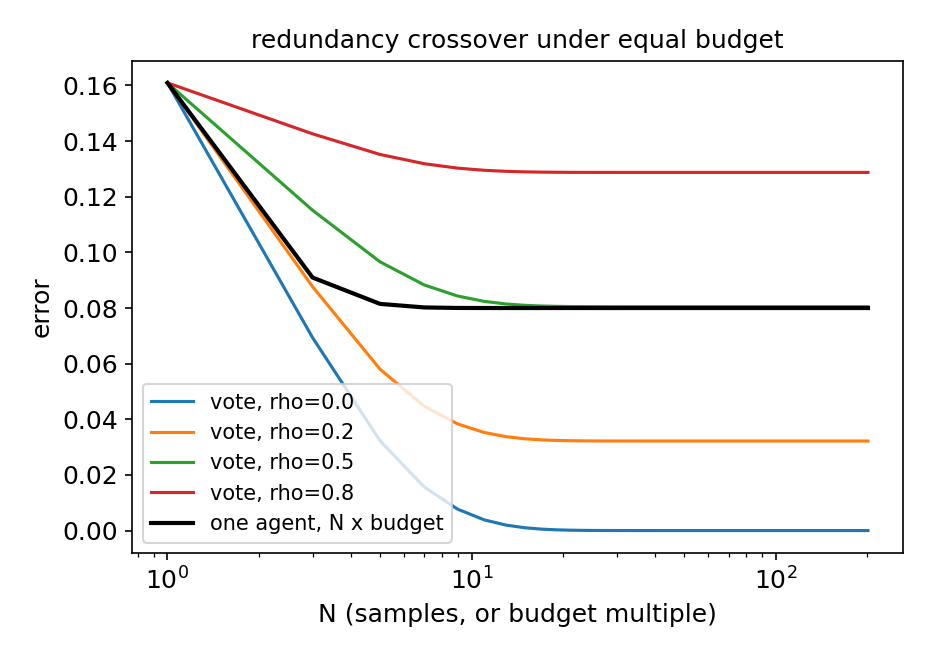}
\caption{Failure probability of a majority vote over $N$ samples at four common-shock correlation levels, against one agent given $N$ times the budget, for a thinking curve $\eps(b)=0.08+0.22e^{-b}$ at $b=1$. The threshold from Theorem~\ref{thm:parseq} is $\rho^{*}=0.08/0.161=0.497$. At every plotted $N>1$, correlations of 0 and 0.2 favour parallel sampling and 0.5 and 0.8 favour sequential thinking.}
\label{fig:redundancy}
\end{figure}

The finite-$N$ comparison goes through the binomial tail in \eqref{eq:vote}. Figure~\ref{fig:redundancy} shows it for the common-shock law with a saturating exponential thinking curve, where $\rho^{*}=0.497$ and the four correlation levels split as the theorem says. Theorem~\ref{thm:parseq} is the equal-budget version of the question \citet{tran2026single} raise and \citet{yang2026diversity} answer without a budget. Debate does not obviously escape it: \citet{choi2025debate} show that debate conserves expected belief under their conditions, which rules out systematic improvement of expected belief over the initial round, though it does not by itself pin the accuracy of thresholded decisions, so we treat the vote as the reference point rather than a proven ceiling. At one step, the ``more agents or more thinking'' question reduces to comparing a shared-failure mass with a thinking floor.

\begin{corollary}[The million-step regime]
\label{cor:maker}
With $m=1$ and per-step redundancy $N$, $-\log S\le L\,[-\log(1-\Evote(N))]+(L-1)h$, with $\Evote$ evaluated at the fresh agents' load $c_0+s+k$ and equality when the first step runs at that load too. For $S$ to stay above a fixed target $S_0>0$ as $L$ grows, both $-\log(1-\Evote(N))$ and $h$ must be $O(1/L)$, so at $L=10^{6}$ the per-step vote failure and the handoff tax must both be of order $10^{-6}$ or smaller. The first is available at minimal context and small shared-failure mass. The second requires the handoff to lose essentially nothing, which is what passing the complete explicit state, rather than a summary, is designed to achieve.
\end{corollary}

This is what \citet{meyerson2025million} do: the object handed between steps is the full disk configuration, so $h$ is near zero by design and the engineering effort goes into driving the vote failure down. The second condition is not optional, since $S\le e^{-(L-1)h}$ regardless of how good the vote is.

\section{Verification}
\label{sec:verify}

Corollary~\ref{cor:verified} says how a verifier enters the crossover: through $q_{\mathrm{eff}}=q(1-\rho_{gv})$, the retry's quality, and the generation budget they displace. What it does not say is how large $q_{\mathrm{eff}}$ is at a boundary versus inside a trajectory, which is an empirical question.

\begin{proposition}[Where verification pays, conditional on a hypothesis]
\label{prop:verify}
Fix the generator's failure $e$, the nominal recall $q$, the retry failure $e_r<1$, and the budget allocation. Suppose $\rho_{gv}$ is smaller for a fresh-context verifier at a decomposition boundary than for a self-check inside the generating trajectory. Then (i) the reduction in per-step failure, $q_{\mathrm{eff}}\,e\,(1-e_r)$, is larger at the boundary, so verification gains accrue disproportionately to decomposed systems, and (ii) with a fixed number of verifiers of equal $q_{\mathrm{eff}}$ and retries as good as the first attempt, the log-loss removed at step $t$ is $\log(1+q_{\mathrm{eff}}e_t)\approx q_{\mathrm{eff}}e_t$, largest at the steps with the largest $e_t$: late in a single-agent trajectory, or at the end of each chunk in a decomposed one.
\end{proposition}

The hypothesis is plausible and not established. It is consistent with the self-correction literature since \citet{huang2024selfcorrect}, which finds that a model asked to reconsider the same evidence in the same context tends to reproduce its error, and with the mechanism in \citet{ao2026reliability}, where a verifier helps only by contributing a signal the generator did not have. A fresh context carries no new exogenous information, but it changes how a bounded agent processes the evidence it has, which is the channel the hypothesis is about. A test of the hypothesis needs the same verifier prompt run in both positions on the same generations, and we list it among the predictions.

The relation to density evolution is an analogy. Replace ``confidently wrong with probability $\eps$'' by ``abstains with probability $\eps$'', take the curve flat, set $\rho=0$, and put a check node at each boundary: the chain becomes a path in the factor-graph language of \citet{aghazadeh2026reliability}, and our one-retry recursion is a one-iteration truncation of their iterative decoding. Their theory does much more on that path (multiple rounds, stopping sets, thresholds) than ours, and ours does two things theirs excludes (context-dependent failure, correlated failure). The two overlap on that path and nowhere else.

\section{Reductions}
\label{sec:reductions}

Table~\ref{tab:reductions} states, for each prior result, the restriction of the model under which its conclusion is recovered and how tight the correspondence is. Three are discussed below.

\begin{table}[t]
\caption{Each prior result as a restriction of the model. \emph{Exact}: the restricted model's theorem has the same statement. \emph{Structural}: the same conclusion under the same premise, reached in a different formal setting. \emph{Analogical}: the mechanism corresponds, but the quantities differ and no reduction is claimed. \emph{Empirical}: a measured result the model reproduces qualitatively.}
\label{tab:reductions}
\centering
\footnotesize
\setlength{\tabcolsep}{2.5pt}
\begin{tabular}{@{}llll@{}}
\toprule
Prior result & Restriction & Recovered & Tightness \\
\midrule
\citet{tran2026single}, DPI & flat curve, equal budget, undegraded & Cor.~\ref{cor:flat} & structural \\
\citet{ao2026reliability}, Prop.~6, Thm.~8 & flat curve, capacity-free centre & Cor.~\ref{cor:flat}, Def.~\ref{def:h} & analogical \\
\citet{tang2025task}, Props.~2.1, 2.2 & flat, $\rho=0$, selector $r$, free compute & Thm.~\ref{thm:vote} per step & structural \\
\citet{su2026when}, Thms.~4.1 to 4.3 & $h=0$ vs.\ $h>0$ on task edges & dependency dichotomy & analogical \\
\citet{yang2026diversity}, Thm.~4.3 & $L=1$, exchangeable samples & saturation at the shared-failure mass & analogical \\
\citet{patel2026sixsigma} & $L=1$, $\rho=0$ & exponential gain, Thm.~\ref{thm:vote} & exact \\
\citet{yang2026auditing}, App.~E.1 & $L=1$, $\rho>0$ & voting fails, Thm.~\ref{thm:vote} & structural \\
\citet{choi2025debate}, \citet{liu2026martingale} & belief-conserving aggregation & vote as reference point & analogical \\
\citet{aghazadeh2026reliability}, Thm.~8 & flat, $\rho=0$, erasure, one retry & one-retry recursion, Eq.~\eqref{eq:retry}, Cor.~\ref{cor:verified} & analogical \\
\citet{xu2025divide} & parallel chunks, no budget, no $\rho$ & task noise as $h$, model noise as curve & analogical \\
\citet{meyerson2025million} & $m=1$, per-step vote, full state passed & Cor.~\ref{cor:maker} & empirical \\
\bottomrule
\end{tabular}
\end{table}

\paragraph{The sceptic corner.}
Under a flat curve, Corollary~\ref{cor:flat} says the single agent weakly dominates at every depth and chunk size, which is the conclusion of \citet{tran2026single} under their premise and of \citet{ao2026reliability} for a centre with no capacity limit. The first is structural: same premise, same conclusion, a bookkeeping argument in place of an information inequality. The second is analogical, because their loss is a Bayes-risk gap under a proper scoring rule and ours is an operational log-reliability, and we do not identify the two.

\paragraph{The optimist corner is a redundancy result.}
The result of \citet{tang2025task} is often cited as a decomposition result. Their single agent has the same per-step success $\sigma(w)$ as each of their workers, so there is no attention tax in their model, and their workers never hand anything off, so there is no handoff tax either. What their multi-agent system has is $N$ independent answers per step and a selector with recall $r$. In our terms that is the redundancy primitive at every step with $\rho=0$, the any-correct aggregator, and no budget constraint. Their unbounded growth in depth is the compounding of a per-step advantage: whenever $r(1-(1-\sigma)^{N})>\sigma$ the ratio of the two success probabilities is $\bigl(r(1-(1-\sigma)^{N})/\sigma\bigr)^{d}$, which grows geometrically in $d$. Their width effect lowers $\sigma=u^{w}$ and so raises the value of redundancy per step. Both survive in our model as long as $\rho$ is small and compute is free. Impose the budget and Theorem~\ref{thm:parseq} takes over. Impose shared failure and the per-step floor becomes $(1-r)+r\rho\eps$ instead of $1-r$. \citet{tang2025task} prove a correct theorem about free, independent redundancy, and it says little about the decomposition question it is often cited for.

\paragraph{The PAC dichotomy is an analogy across phases.}
\citet{su2026when} find that multi-agent fine-tuning is more sample-efficient when subtasks are independent (Theorem~4.3 there) and worse when they are dependent (Theorem~4.2, a $K^{2}$ penalty from error propagation). In our inference-time model a dependency edge is an edge where the downstream step needs information from upstream, which is where $h>0$, and an independent edge is where $h=0$. We record it as a correspondence across a change of phase and claim nothing more.

\section{Two contradictions, read through the model}
\label{sec:contradiction}

The literature contains two separate contradictions, and they resolve through different parameters.

\subsection{Redundancy: free independent samples against budgeted dependent ones}

Claim A$_{\mathrm{red}}$ (\citealp{tang2025task}, \citealp{patel2026sixsigma}): the gain from $N$ agents per step is exponential in $N$ and compounds without bound in depth. Claim B$_{\mathrm{red}}$ (\citealp{tran2026single}, \citealp{yang2026diversity}, \citealp{choi2025debate}): at equal budget more agents buy no more than more thinking, homogeneous scaling saturates, and debate does not improve on its first round. In the model, Claim A$_{\mathrm{red}}$ is Theorem~\ref{thm:vote} at $\rho=0$ with unconstrained compute, and Claim B$_{\mathrm{red}}$ is Theorem~\ref{thm:parseq} on the sequential side of the threshold. Neither claim holds at every level of dependence. Parallel sampling wins at equal budget below the threshold and loses above it, and the empirical success of self-consistency \citep{wang2023selfconsistency} says the shared-failure mass for a well-prompted sampled chain is often below threshold on the tasks people try. Context plays no part in this contradiction. What decides it is who paid for the samples and whether they failed together.

\subsection{Decomposition: a capacity-free centre against a degrading one}

Claim A$_{\mathrm{dec}}$ (\citealp{tran2026single}, \citealp{ao2026reliability}): under equal information and equal budget, a single agent that uses its context fully weakly dominates any delegated network. Claim B$_{\mathrm{dec}}$ (\citealp{meyerson2025million}, the expert-designed row of \citealp{jwalapuram2026illusion}): decomposition wins, and wins more on longer and more context-heavy tasks. In the model, Claim A$_{\mathrm{dec}}$ holds for every chunk size under a flat curve (Corollary~\ref{cor:flat}), and more generally whenever the total avoidable tax satisfies $A_\infty\le h$ (Theorem~\ref{thm:chunks}). Claim B$_{\mathrm{dec}}$ holds, for single-step agents, if and only if $a_\infty>h$ (Theorem~\ref{thm:crossover}), and for some chunk size if and only if $A_\infty>h$. For any curve that rises above the fresh agent's operating point by more, in total, than a handoff costs, decomposition wins at every sufficiently large depth and not below its first crossover. For any curve that does not, it never wins. The two camps were proving theorems about different curves and reporting them as theorems about agents.

What each side left out is easy to list once the model is in front of you. \citet{tran2026single} prove dominance for an undegraded centre and, when they degrade it, use a corruption knob rather than accumulated load, so their crossovers are observed but not located from measurements. \citet{ao2026reliability} give the centre no capacity and describe their model as operating above the level of token generation. \citet{tang2025task} have no handoff and no budget. The redundancy papers hold $L=1$ and so cannot speak to decomposition. \citet{aghazadeh2026reliability} exclude the curve and correlation in writing and have no single-agent baseline to compare against.

\subsection{Regimes in the published record}

\begin{figure}[t]
\centering
\includegraphics[width=0.66\linewidth]{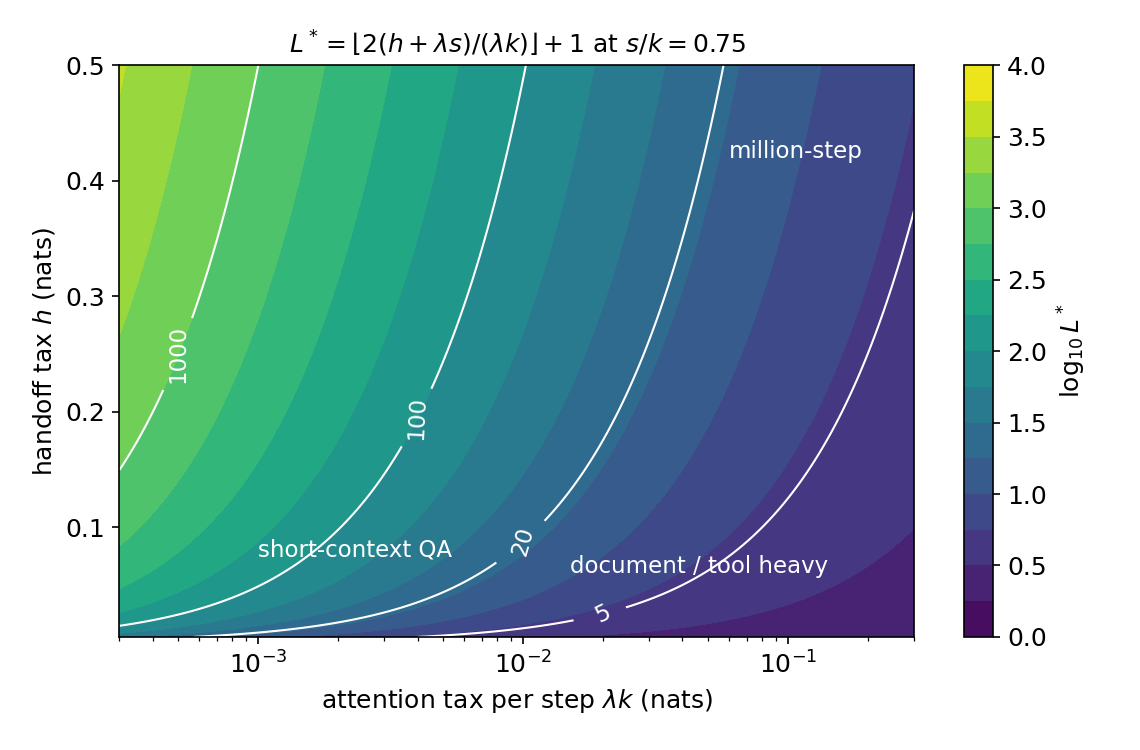}
\caption{Crossover depth $L^{*}=\lfloor 2(h+\lambda s)/(\lambda k)\rfloor+1$ over the attention tax per step $\lambda k$ and the handoff tax $h$, at $s/k=0.75$. White contours mark $L^{*}\in\{5,20,100,1000\}$. The three annotations are qualitative placements of the benchmark families discussed in the text, not fitted points.}
\label{fig:regime}
\end{figure}

The model says the winner is decided by where $(c_0,k)$ sits on the curve and by $h$. Figure~\ref{fig:regime} draws $L^{*}$ over $(\lambda k,h)$ and places three families of published results qualitatively. Placement is qualitative, since scaffold choice moves scores as much as model choice does \citep{zhang2026harness}.

In the flat region, with small $k$ and short contexts, $a_t$ stays near $a_1$ and any positive $h$ makes decomposition lose. This is multi-hop question answering and the static reasoning benchmarks. \citet{tran2026single} find single agents match or beat multi-agent systems there at matched tokens, and \citet{jwalapuram2026illusion} find self-consistency ahead of six automated frameworks on reasoning benchmarks such as GPQA-Diamond and HLE-Maths, often at under a tenth of their cost. Redundancy still helps in this region through Theorem~\ref{thm:vote}, and self-consistency rather than a plain chain of thought wins in that audit for that reason.

In the rising region, with large $k$ or a large $c_0$ and a decomposable task, the same audit gives the opposite answer. Their SMFR task embeds investor transactions in long price tables, with parallel investors and sequential dependence within each investor. An expert-designed decomposition with a meta-agent, per-investor extractor and calculator agents, and deterministic orchestration takes GPT-5 from 57.0 with self-consistency to 96.5, at a cost of 554.82 against 478.40 dollars for the baseline, and takes GPT-OSS-120B from 26.1 to 36.1. That is the direction Theorem~\ref{thm:crossover} predicts: fresh contexts per investor remove the attention tax the single agent pays on the full table. The automated frameworks gain little on the same task, at most 6.6 points and at roughly two to five times the cost, and the authors' own deconstruction finds many of them collapsing into ensembling rather than decomposing, so they pay orchestration overhead without buying much reduction in attention tax. We do not place SMFR quantitatively on Figure~\ref{fig:regime}: the source describes the task's context in more than one way and we have no curve for GPT-5.

The same table contains the competency floor. GPT-4o moves from 22.1 to 22.6 under the expert decomposition, which is nothing. In the model, if $\eps$ at minimal context is already large, decomposition removes the attention tax but the fresh agents' product $\exp(-L\,g(c_0+s+k))$ is still near zero, because nothing in a handoff lowers a failure rate that is already high at short context.

Our own pilot in Section~\ref{sec:calib} places a 20B model in the flat region, to within its resolution, for arithmetic failures on the whole measured range, and in a rising region for failures of instruction compliance, with the knee poorly located. The ledger experiment of Section~\ref{sec:exp} places the same model in the rising region on a task where every step depends on the trajectory. In the steep, long region there is \citet{meyerson2025million}, where no single agent survives $10^{6}$ steps and $m=1$ with per-step voting and full-state handoffs gives Corollary~\ref{cor:maker}.

\subsection{What would falsify this}

Six predictions, roughly in order of how cheap they are to test. Section~\ref{sec:exp} tests the first on one task and one model, on the winning side of the crossover.

The crossover exists and moves. Run the equal-budget comparison of \citet{tran2026single} on tasks with larger per-step context growth $k$, such as tool-heavy or document-heavy multi-hop, with a measured handoff tax, and the winner flips at a depth predicted by \eqref{eq:lstar}.

$L^{*}$ moves with the model being served. For a fixed task, harness, failure definition, and measured $h$, models with flatter measured curves have larger $L^{*}$, and the single-agent window widens with each generation of long-context improvement.

Chunk size. For a long decomposable task with measured $(\lambda,k,h)$, agents owning about $\sqrt{2h/(\lambda k)}$ steps beat both one-step-per-agent and one-agent-for-all designs.

Summary length. Performance as a function of summary length has an interior optimum where $-h'(s)=\lambda$, so fixed-ratio compaction is off-peak for some models.

Redundancy. The parallel-versus-sequential winner at equal budget is predicted by the shared-failure mass against $\eps_\infty$. Heterogeneous ensembles flip more tasks to parallel and homogeneous self-consistency saturates, as \citet{yang2026diversity} report.

Verification. The same verifier prompt reduces failure by more at a decomposition boundary than as a self-check inside one trajectory, and the gap grows with context load.

\section{An exploratory pilot}
\label{sec:calib}

This pilot measures one axis of the model, the context degradation curve, on one model and task family. Section~\ref{sec:exp} provides the architecture-level test.

\paragraph{Protocol.}
The model is gpt-oss-20b as packaged by Ollama (tag \texttt{gpt-oss:20b}, 20.9B parameters, MXFP4 quantisation, native context 131,072), served through Ollama 0.30.8 on one RTX~3090 with a derived model fixing the context window at 71,680 tokens (\texttt{num\_ctx}), flash attention, an 8-bit KV cache, the reasoning level fixed at low through the harmony system message, sampling temperature 0.7 in place of the packaged default of 1, and the output budget fixed at 1,024 tokens including reasoning, which together fix $b$. The task is a family of chained arithmetic word problems with six operations and an exact integer answer, generated from a seed, the same 100 problems at every load so that difficulty is matched across loads. Context load is varied by prepending well-formed, digit-free prose presented as earlier session material, at nominal 0, 2k, 8k, 32k, and 64k tokens, which is the protocol of \citet{du2025contextlength} pointed at one model. What this measures is length-induced degradation under padded session context, with the instruction at one end and the question at the other, and not the distribution of material a real agent trajectory accumulates. The system prompt, placed at the top, asks for a final line of the form \texttt{ANSWER: <integer>}. Each problem is sampled twice at temperature~0.7, 1,000 calls in all. Measured loads by the o200k tokeniser are 107, 2,117, 8,023, 31,641, and 63,129 tokens. Only the final 200 characters of each response were stored, so finish reasons and token counts are not available for the pilot. The generator, filler, grader, and stored responses are in the repository.

\begin{figure}[t]
\centering
\includegraphics[width=0.98\linewidth]{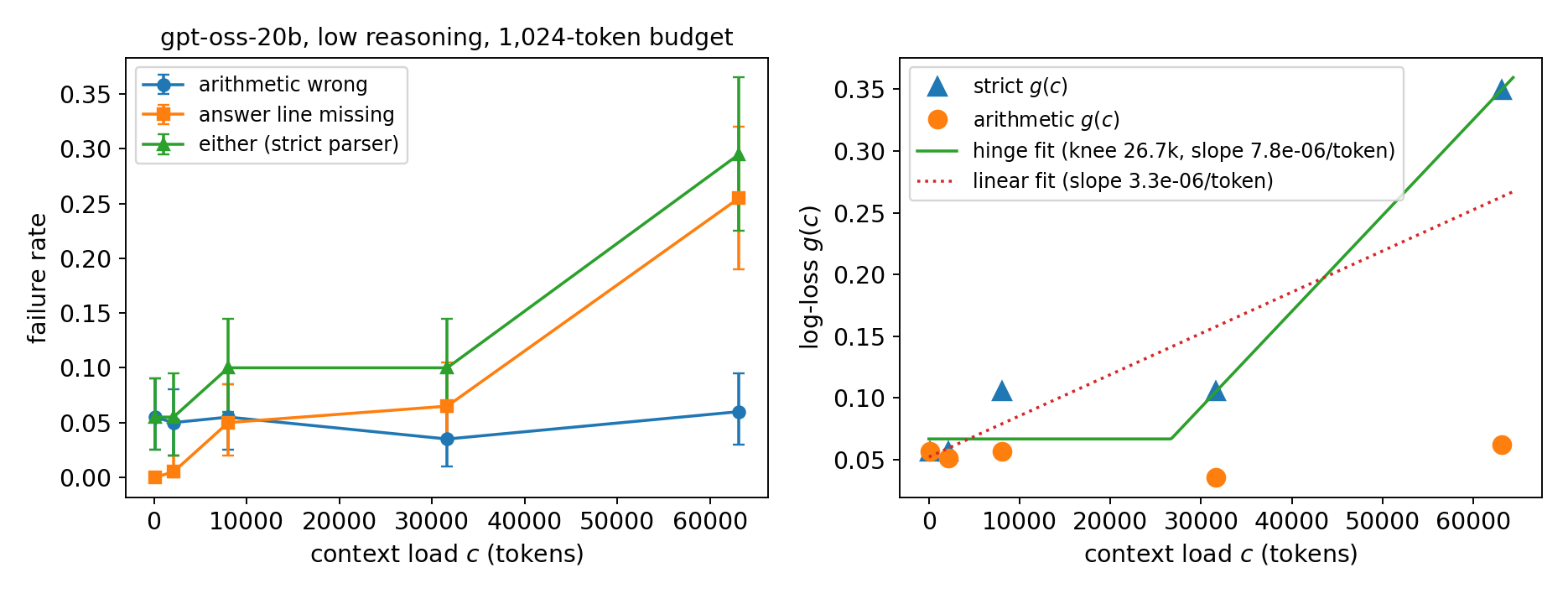}
\caption{Left: the three failure definitions against context load, with 95\% cluster-bootstrap intervals over problems. Right: log-loss for the arithmetic and strict definitions, with a weighted hinge fit to the strict curve and the linear fit for comparison. The arithmetic curve is flat to within measurement. The strict curve is flat, then not, with the location of the change poorly determined.}
\label{fig:calibration}
\end{figure}

\paragraph{Two curves from one run.}
Table~\ref{tab:calibration} and Figure~\ref{fig:calibration} show what came back, with intervals from a cluster bootstrap over problems, since the 200 calls at a load are two samples on each of the same 100 problems. Graded on the arithmetic alone, the failure rate is 5.5, 5.0, 5.5, 3.5, and 6.0 per cent across the five loads. A paired comparison by problem of the 64k load against the other four gives a difference of $+1.1$ points with standard error 1.9 ($t_{99}=0.59$), and the cluster-bootstrap interval on the weighted slope of $g$ is $[-6.1,+6.6]\times10^{-7}$ per token, which over the measured range permits a change of about four percentage points in either direction. The arithmetic curve is flat to within four points over 63k tokens at this budget. Graded on whether the response contained the requested answer line, sitting 63k tokens upstream at the top load, the picture is the opposite: 0, 0.5, 5.0, 6.5, and 25.5 per cent of responses omit it. The paired-by-problem difference for the 64k load is $+22.5$ points with standard error 3.4 ($t_{99}=6.7$), and the failures are spread across problems (33 problems fail once, 9 twice, 58 never). In 47 of the 51 the number was still right. The model just stopped saying so in the form it was asked to, and a strict parser downstream fails on all 51. The stored tails at 64k show no obvious mid-response truncation. The strict curve, counting a step as failed if either the number or the format is wrong, is 5.5, 5.5, 10.0, 10.0, and 29.5 per cent, and at 64k its 59 failures fall on 47 problems (35 once, 12 twice).

\begin{table}[t]
\caption{Pilot on gpt-oss-20b at low reasoning and a 1,024-token budget, 200 calls per load (100 problems, two samples). Arithmetic: wrong final number. Format: required answer line missing. Strict: either. Intervals are 95\% cluster-bootstrap intervals over problems.}
\label{tab:calibration}
\centering
\small
\begin{tabular}{@{}rrrrr@{}}
\toprule
$c$ (tokens) & arithmetic failures & format failures & strict failures & strict $\eps(c)$ [95\% CI] \\
\midrule
107 & 11 & 0 & 11 & 0.055 [0.025, 0.090] \\
2,117 & 10 & 1 & 11 & 0.055 [0.020, 0.095] \\
8,023 & 11 & 10 & 20 & 0.100 [0.060, 0.145] \\
31,641 & 7 & 13 & 20 & 0.100 [0.060, 0.145] \\
63,129 & 12 & 51 & 59 & 0.295 [0.225, 0.365] \\
\bottomrule
\end{tabular}
\end{table}

\paragraph{The fit.}
A weighted hinge fit, searching knee locations from 0 to 32k tokens, places the knee near 27k and gives a post-knee slope of $7.8\times10^{-6}$ per token, with a 95\% cluster-bootstrap interval of $[3.4,11.1]\times10^{-6}$. The knee location is loosely determined. The robust feature is the rise from roughly 10 per cent strict failure at 32k to roughly 30 per cent at 63k.

\paragraph{Implications of the pilot.}
Under the arithmetic definition no slope is detectable, so to the pilot's resolution this model sits in the flat corner of Corollary~\ref{cor:flat} for trajectories under 64k tokens, where decomposition has no attention tax to buy back. Under the strict definition the hinge fit and Proposition~\ref{prop:knee} give, for assumed handoff costs of $h=0.02$ and $h=0.05$ nats and no summary overhead: for a step adding $k=400$ tokens, the ledger's step size, the single agent reaches the fitted knee at step 67 and decomposition pays from depth 103 or 131. For a tool-heavy step adding $k=2{,}000$ tokens the knee arrives at step 14 and the crossover at depth 20 or 26. These are brute-force evaluations of $D$, and the closed form is within one step of each. Under a knee-shaped curve the chunk size is not Theorem~\ref{thm:chunk}'s square root, because below the knee a handoff buys nothing: agents should own as many steps as fit under the knee, and only past that does the square-root trade-off apply. These crossover depths are illustrative because they use the fitted knee and assumed $h$. Section~\ref{sec:exp} performs the architecture-level test with measured quantities.

\paragraph{Failure definitions.}
The curve is a property of the model, the task, and the failure definition together. The same thousand calls give a flat curve and a steep one, and a naive last-integer grader produced a third by treating a remainder in responses without an answer line as the final answer. The mechanism in the steep case is not reasoning collapse but decay of compliance with an instruction that sits far from the point of use, which is the phenomenon \citet{liu2024lost} measured for facts, seen here for an instruction. Two consequences follow if the mechanism is what it appears to be. A practitioner who repeats the format instruction next to the question should flatten the strict curve, and the model predicts this because it shortens the distance the instruction must survive. And a paper that reports a single-agent win on short-context benchmarks and a multi-agent win on long-context ones may be measuring exactly this: not what the model can compute, but what it still remembers it was asked to do.

\section{The experiment}
\label{sec:exp}

The model suggests one direct test. Measure the curve and the handoff tax, compute the depth at which decomposition should start to win and how often each system should succeed, then compare with both systems run on the same ledgers. We ran that test on the ledger task of the running example.

\paragraph{Task.}
A ledger of $L$ transactions with an exact integer running balance, starting at 50,000. Six transaction kinds: deposit, withdrawal, a percentage fee on the current balance rounded down, a hold that removes an amount now, a release that returns a hold's amount three to twelve steps later but names the hold only by id, and a reversal that undoes a deposit or withdrawal three to twelve steps back, also named only by id. The last two are the long-range dependencies: the amount has to come from earlier context or from a handoff note. Each transaction block is padded with digit-free administrative memo lines to about 400 tokens, and the measured context growth was 398 tokens per step. Ledgers are generated from a seed, and the same 100 ledgers are used at each depth by both systems, so every comparison is paired. A step is graded on its increment, whether the reported balance moved by the true amount for that transaction, so a step is not penalised for a balance error it inherited. A ledger is graded twice: on whether every non-fee increment is right and every handoff note carries its balance forward unchanged (the stationary grading, explained below) and on whether its final balance is exactly right.

\paragraph{Systems and protocol.}
The single agent runs one growing conversation, one call per transaction, so its context at step $t$ is $c_0+kt$ with $c_0=197$ tokens. The decomposed system gives each agent $m=5$ consecutive transactions in a fresh conversation. At each boundary the outgoing agent is asked for a handoff note of about 300 tokens (225 words) carrying the balance, every open hold, and the id and amount of the last twelve transactions, and the incoming agent starts from that note. The handoff tax was measured on a separate pool of 100 ledgers by the paired design of Definition~\ref{def:h}: twelve transactions run once, then the next five run twice, continuing in the same conversation and starting from a note, at note allowances of 100, 300, and 900 tokens (produced notes 137, 183, and 255 tokens). The model is gpt-oss-20b on a hosted OpenAI-compatible endpoint (provider recorded in the run configuration), at reasoning level low, temperature 0.7, and a 512-token allowance per step call including reasoning, which is $b$, the same for every step call in both systems. Handoff notes were separate calls with a 600-token allowance, so the writing agent's own step budget was not reduced: the experiment sits at $\Delta_o=0$ in the accounting of Section~\ref{sec:model}, and the note tokens are overhead on top of matched step budgets. Actual generation per ledger, reasoning plus answer, averaged 806, 2,873, and 10,046 tokens for the single agent at depths 20, 50, and 100, against 1,611, 4,797, and 10,195 for the decomposed system including its notes (770, 2,001, and 4,078 for its step calls alone). Per step call the decomposed agents used 39 to 41 tokens at every depth while the single agent's use rose from 40 to 100 with load, so at depth 100 the two systems generated the same number of tokens and at shallower depths the decomposed system generated more, almost all of it in notes. On the input side the single agent processed 2.0M prompt tokens per ledger at depth 100 against 0.2M, since it re-reads its growing context at every step. It is the same model as in the pilot, on a different serving stack, so the two sections' curves are not compared quantitatively. In all, 40,600 calls, 309M prompt tokens, about ten dollars, and 27 minutes of wall-clock time. Finish reasons and token counts were kept for every call.

\paragraph{Analysis.}
The curve is interpolated through its bins, fee steps are left out for the reason given below, and each ledger's prediction counts the steps that are actually scored, at their actual positions. Every numerical input (curve, handoff tax, $k$, $c_0$, and note length) comes from single-agent and handoff runs. These choices were fixed once all runs were complete, and an earlier hinge-based analysis is kept in the repository.

\begin{figure}[t]
\centering
\includegraphics[width=0.98\linewidth]{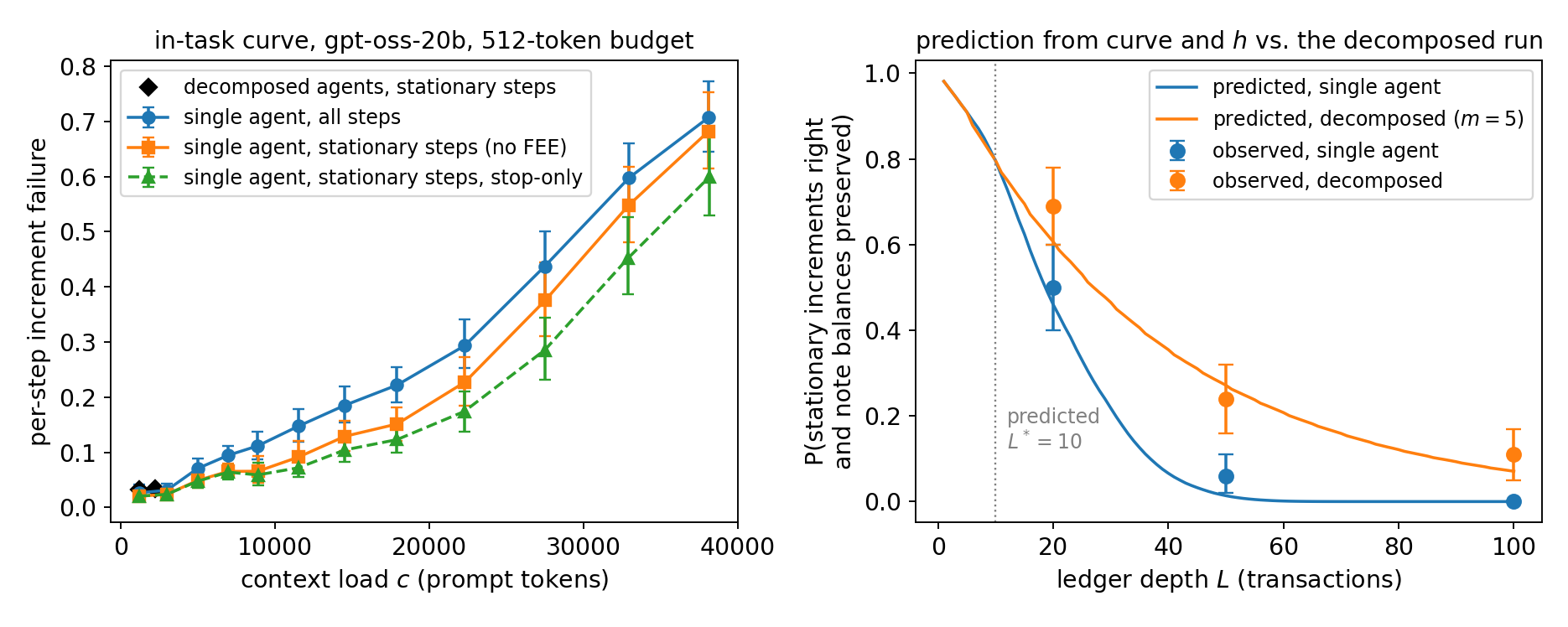}
\caption{Left: the in-task curve from single-agent steps, on all kinds, on the stationary kinds, and on stationary kinds with truncated calls removed, with the decomposed agents' stationary steps at their own loads. Intervals are cluster-bootstrap over ledgers. Right: predicted success against depth for both systems from the curve and the measured handoff tax, evaluated on prefixes of the depth-100 ledgers, with the observed rates at depths 20, 50, and 100 and bootstrap intervals over ledgers. The dotted line is the predicted crossover.}
\label{fig:exp}
\end{figure}

\paragraph{The curve.}
Table~\ref{tab:expcurve} and the left panel of Figure~\ref{fig:exp} show per-step failure against context load from the 17,000 single-agent steps. It rises from 2.7 per cent below 2k tokens to 71 per cent above 36k, far steeper than in the pilot. Unlike the pilot's irrelevant padding, every step here depends on the accumulated trajectory. Two effects are separated from the load effect. First, the model's reasoning grew with load, from 20 tokens per step below 4k to 153 above 30k, and 5.1 per cent of single-agent calls hit the 512-token budget against 0.13 per cent of decomposed calls. Those count as failures under the fixed per-call budget $b$. Since they are asymmetric, the curve is also reported on calls that finished normally and the depth comparison on ledgers with no truncated call, and both preserve the direction at every depth. Second, fee steps drift with position for a reason unrelated to load. A fee is a percentage of the incoming balance, so once a balance is wrong every later fee is graded wrong even when its arithmetic is right. In the decomposed system, at constant load, fee failure rises from 7 per cent in the first ten transactions to 72 per cent after the fiftieth. Conditioned on a correct preceding balance it is flat, 5 of 91 early and 3 of 55 late, while 283 of 340 late fees that follow a wrong balance fail. Fee steps therefore measure the state carried into them rather than the step itself, and the curve, the handoff tax, and the prediction use the other five kinds (the stationary steps). A ledger then has 18.0, 45.0, and 90.0 scored steps on average at depths 20, 50, and 100, and the full-task result is reported alongside.

\begin{table}[t]
\caption{The in-task curve from single-agent steps on the ledger task, gpt-oss-20b at a 512-token budget. Failure is the per-step increment failure rate. The stationary column excludes fee steps, and stop-only excludes calls that hit the budget. Seven of the twelve load bins, which together hold 9,178 of the 17,000 steps, are shown. The prediction and Figure~\ref{fig:exp} use all twelve. Intervals are cluster-bootstrap over ledgers.}
\label{tab:expcurve}
\centering
\small
\begin{tabular}{@{}rrrrr@{}}
\toprule
load (tokens) & steps & all kinds [95\% CI] & stationary kinds & stationary, stop-only \\
\midrule
0 to 2k & 1,200 & 0.027 [0.015, 0.042] & 0.021 & 0.021 \\
4k to 6k & 1,500 & 0.071 [0.055, 0.089] & 0.049 & 0.048 \\
8k to 10k & 1,100 & 0.112 [0.087, 0.138] & 0.066 & 0.059 \\
13k to 16k & 1,403 & 0.185 [0.154, 0.219] & 0.129 & 0.104 \\
20k to 25k & 1,393 & 0.294 [0.253, 0.341] & 0.227 & 0.174 \\
30k to 36k & 1,509 & 0.598 [0.543, 0.661] & 0.548 & 0.452 \\
36k to 45k & 1,073 & 0.707 [0.646, 0.773] & 0.682 & 0.599 \\
\bottomrule
\end{tabular}
\end{table}

The decomposed agents sit close to the same curve. Their stationary steps, at loads of 1.2k and 2.2k tokens, fail at 3.3 and 3.4 per cent against about 2 per cent for the single agent at the same loads.

\paragraph{The handoff tax.}
Continuing from history, at a mean load of 6.2k tokens, the post-boundary stationary steps failed at 5.2 per cent. Starting from a note, at a mean load of 1.6k, they failed at 3.2, 2.0, and 1.8 per cent for the three note lengths. The raw comparison therefore favours the note, because the history sits higher on the curve, and Definition~\ref{def:h} adds the curve term back, summed over each chunk's scored steps at their observed loads: $0.172$ nats per chunk. This gives an increment tax $h_{\mathrm{inc}}$ of $0.077$, $0.026$, and $0.015$ nats for notes of 137, 183, and 255 tokens.

The stationary endpoint also requires each note to preserve the incoming reported balance. The same parser grades this event in calibration and evaluation. A missing or unparseable balance fails. Preservation succeeds on 100, 99, and 100 of the 100 calibration ledgers. Under the product approximation, its separate boundary cost is $h_{\mathrm{bal}}=-\log q_{\mathrm{bal}}$, so the total reference-chunk taxes are $0.077$, $0.036$, and $0.015$ nats, still nonincreasing in note length. At the 300-token setting $h_{\mathrm{bal}}=-\log(0.99)=0.010$. A bootstrap over single-agent and paired handoff ledgers re-estimates the curve, increment tax and balance preservation together. The increment tax has a 95\% interval of $[-0.11,0.16]$ nats. The total tax interval is $[-0.10,0.18]$ nats. Following Assumption~\ref{as:handoff}, forecasts project the increment tax to zero when negative and retain the nonnegative balance cost.

As a sensitivity check, direct whole-chunk success is 0.81 from history and 0.90 from the note, requiring both correct non-fee increments and preserved balance. This gives $h_{\mathrm{joint}}=0.067$. Using the primary total $h=0.036$ and the low-load slope of about $8\times10^{-6}$ per token, Theorem~\ref{thm:chunk}'s reference-chunk approximation gives $\sqrt{2h/(\lambda k)}\approx4.8$ steps, close to the $m=5$ chosen before measuring anything.

\paragraph{Prediction and result.}
For each ledger the prediction sums the interpolated curve over its scored steps, at loads $c_0+kt$ for the single agent and at the fresh-context loads of their chunk positions for the decomposed system, with fee steps still adding context. The increment tax $h_{\mathrm{inc}}=0.026$ is spread over the calibration chunk's 4.41 scored steps on average, on every scored step after the first chunk. The balance cost $h_{\mathrm{bal}}=0.010$ is paid once per actual handoff, including a handoff followed only by fee steps. This extends product form to both increment and balance-preservation events. Predicted success is $\exp(-G)$ averaged over the 100 ledgers. On prefixes of the depth-100 ledgers this puts the crossover at $L^{*}=10$, with bootstrap quantiles of 6 and 27, and gives the rates in Table~\ref{tab:expresult} and Figure~\ref{fig:exp}.

Decomposition won at every tested depth on both gradings. The paired improvements in stationary success are 0.19, 0.18, and 0.11, with bootstrap intervals $[0.05,0.33]$, $[0.08,0.28]$, and $[0.05,0.18]$, and exact McNemar $p$ of 0.014, 0.001, and below 0.001 (0.049, 0.003, and 0.002 on the final balance). The predictions miss the observations by at most 9 percentage points. For the single agent the observed rate is 4, 5, and 0 points above the prediction, the direction product form implies when a single agent's failures cluster on hard ledgers. For the decomposed system the misses are $+8.4$, $-3.1$, and $+3.9$ points, with no consistent sign. The ordering does not depend on the estimator: replacing the increment tax with $h_{\mathrm{joint}}-h_{\mathrm{bal}}$ while retaining the once-per-boundary balance cost gives decomposed predictions of 0.55, 0.20, and 0.04, still ahead of the single agent at every depth. Restricting to ledgers on which neither system hit the token limit leaves 98, 62, and 12 pairs, with final-balance success of 0.49 against 0.65, 0.02 against 0.16, and 0 against 0.08.

\begin{table}[t]
\caption{Predicted and observed success by depth on the ledger task, 100 paired ledgers per depth. Stationary: every non-fee increment right and every note's balance carried forward. Final: final balance exactly right. Predictions use the stationary curve, increment tax $h_{\mathrm{inc}}=0.026$, balance cost $h_{\mathrm{bal}}=0.010$, and each ledger's scored steps and boundaries, with every input taken from single-agent and handoff runs. Paired $p$ values are exact McNemar tests.}
\label{tab:expresult}
\centering
\small
\begin{tabular}{@{}rrrrrrrr@{}}
\toprule
 & \multicolumn{2}{c}{predicted, stationary} & \multicolumn{2}{c}{observed, stationary} & \multicolumn{2}{c}{observed, final balance} & \\
\cmidrule(lr){2-3}\cmidrule(lr){4-5}\cmidrule(lr){6-7}
$L$ & single & decomposed & single & decomposed & single & decomposed & paired $p$ (stationary / final) \\
\midrule
20 & 0.46 & 0.61 & 0.50 & 0.69 & 0.49 & 0.64 & 0.014 / 0.049 \\
50 & 0.01 & 0.27 & 0.06 & 0.24 & 0.04 & 0.18 & 0.001 / 0.003 \\
100 & 0.00 & 0.07 & 0.00 & 0.11 & 0.00 & 0.10 & $<$0.001 / 0.002 \\
\bottomrule
\end{tabular}
\end{table}

\paragraph{Scope.}
The measured curve and handoff tax predicted the winning architecture at all three tested depths, and the decomposed system's success to within 9 percentage points. All three sit past the predicted crossover at depth 10, and runs at depths either side of it are the natural next measurement.

\section{Scope and extensions}
\label{sec:limits}

The model is first order in three places. Product form ignores dependence along a trajectory. When failures cluster on hard instances it overstates the single agent's loss, which is the direction seen on the ledger, and a dependent-failure version of the crossover is a natural extension. The budget is fixed per step rather than in total, and the ledger experiment matches systems the same way. The handoff tax is one scalar per boundary, related to the information-theoretic loss of \citet{ao2026reliability} by analogy. The redundancy floor depends on the whole mixing law, of which the common-shock law is one example.

The measurements cover one model on two tasks. The quantities the model asks for next are the boundary verifier recall of Section~\ref{sec:verify}, the thinking floor of Section~\ref{sec:redundancy}, and curves for further models, each of which turns into a direct test of a result above.

\section{Conclusion}

Inside the model, the single-versus-multi-agent question turns on one curve and one tax rather than on the number of agents. Decomposition trades information for attention: it pays a handoff tax at every boundary to buy back per-step accuracy, and it is worth it when the tax a fresh agent avoids over its run exceeds what the handoff costs, net of what fresh-context verifiers catch. Redundancy trades budget for independence: it eventually wins at equal compute whenever the mass the samples place on shared failure lies below the thinking floor of one agent. The formal results we could find in the 2025 to 2026 literature correspond to these two statements with some quantities pinned, exactly in a few cases and by analogy in most, and the empirical record, including the one audit that reports both outcomes side by side, sorts along the same lines. The failure definition is part of the curve: the pilot curve was flat for the task and steep for the instruction. On the ledger task, measuring the context curve and handoff tax from single-agent and handoff runs alone was enough to predict the winning architecture and the decomposed system's success to within 9 percentage points, with the crossover itself below the tested range. This turns the framework from a retrospective explanation of conflicting results into a measurable architecture-selection rule in the regime tested here.

\appendix

\section{Proofs}
\label{app:proofs}

\subsection*{Theorem~\ref{thm:crossover}}
From \eqref{eq:D}, the increments of $D$ are $D(L+1)-D(L)=a_{L+1}-h_{\mathrm{eff}}$, nondecreasing in $L$ by Assumption~\ref{as:curve}, so $D$ is discretely convex, and $D(1)=0$.

(ii) If $a_\infty\le h_{\mathrm{eff}}$ then every increment is at most zero, so $D$ is nonincreasing from zero and $D(L)\le 0$ for all $L$.

(i) If $a_\infty>h_{\mathrm{eff}}$, choose $\eta>0$ and $T$ with $a_t\ge h_{\mathrm{eff}}+\eta$ for all $t\ge T$. Then for $L>T$, $D(L)\ge D(T)+(L-T)\eta\to+\infty$, so $D(L)>0$ for all large $L$. Because increments are nondecreasing, once an increment is positive all later ones are, so from the first index at which $D>0$ the sequence stays positive. With $D(1)=0$ the set $\{L:D(L)>0\}$ is an up-set $\{L\ge L^{*}\}$, which gives existence and uniqueness of $L^{*}$.

(iii) $D$ is pointwise decreasing in $h_{\mathrm{eff}}$, so its zero-crossing moves right. Increasing $s$ raises $g(c_0+s+k,b)$ and lowers every $a_t$, again lowering $D$ pointwise. A change to the curve that leaves $[0,c_0+s+k]$ unchanged and raises the curve above it leaves $g(c_0+s+k,b)$ fixed and raises $g(c_0+kt,b)$ for every $t$ with $c_0+kt>c_0+s+k$, so it raises $a_t$ for those $t$ and leaves the others unchanged, raising $D$ pointwise. \qed

\subsection*{Corollary~\ref{cor:flat}}
Under a flat curve $g(c,b)=\gamma(b)$, every step in \eqref{eq:gd} contributes $\gamma(b)$ and there are $\lceil L/m\rceil-1$ handoffs, so $\GD(L,m)=L\gamma(b)+(\lceil L/m\rceil-1)h_{\mathrm{eff}}\ge L\gamma(b)=\GSA(L)$, since $h_{\mathrm{eff}}=h+\Delta_o\ge 0$. This includes a shorter final chunk. \qed

\subsection*{Theorem~\ref{thm:chunks}}
Write $\Sigma_j:=\sum_{i\le j}[g_{\max}-g(c_0+s+ki)]$ and $R(L):=\sum_{t\le L}[g_{\max}-g(c_0+kt)]\ge 0$, the single agent's own distance from saturation, which is nondecreasing in $L$, with $\Sigma_0=0$. For $L=nm+\bar m$, $0\le\bar m<m$, substituting $\sum_{t\le L}g(c_0+kt)=Lg_{\max}-R(L)$ and the corresponding sums into \eqref{eq:gd} with $o=0$ gives
\[
\GSA(L)-\GD(L,m)=(n-1)\bigl(\Sigma_m-h\bigr)+\mathbf{1}_{\{\bar m>0\}}\bigl(\Sigma_{\bar m}-h\bigr)-\bigl[R(L)-R(m)\bigr].
\]
The bracket is nonnegative, and $\Sigma_{\bar m}\le\Sigma_m$. If $\Sigma_m\le h$ every term is nonpositive, so chunk size $m$ never strictly wins, which gives the ``only if''. If $\Sigma_m>h$ the first term grows linearly in $L$, the second takes finitely many values for fixed $m$, and $R(L)=o(L)$ because $g(c_0+kt)\to g_{\max}$, so the difference is positive for all large $L$. Since $\Sigma_m$ is nondecreasing in $m$ with limit $A_\infty$, some $m$ has $\Sigma_m>h$ iff $A_\infty>h$. If $g$ is unbounded, monotonicity and $k>0$ give $\GSA(L)/L\to\infty$, while for fixed $m$ the decomposed loads are bounded and $\GD(L,m)=O(L)$, so decomposition wins at every $m$ for large $L$. \qed

\subsection*{Theorem~\ref{thm:lstar}}
With $o=0$, substituting $a_t=\lambda(k(t-1)-s)$ into \eqref{eq:D} gives $\sum_{t=2}^{L}a_t=\lambda k\,\tfrac{L(L-1)}{2}-\lambda s(L-1)$, hence \eqref{eq:Dlinear}. For $L\ge 2$ the sign of $D(L)$ is the sign of $\tfrac12\lambda kL-\lambda s-h$, which is positive iff $L>2(h+\lambda s)/(\lambda k)$. The smallest integer $L\ge 2$ satisfying a strict inequality $L>x$ is $\max\{2,\lfloor x\rfloor+1\}$. \qed

\subsection*{Proposition~\ref{prop:knee}}
Below the knee $a_t=0$ and above it $a_t=\lambda_2(c_0+kt-c_\kappa)=\lambda_2k(t-T_\kappa)$. Treating $t$ as continuous, $\sum_{t\le L}a_t\approx\tfrac12\lambda_2k(L-T_\kappa)^2$ for $L>T_\kappa$. Setting this greater than $(L-1)h\approx(L-T_\kappa+T_\kappa)h$ and solving the quadratic in $x=L-T_\kappa$ gives \eqref{eq:knee}. \qed

\subsection*{Theorem~\ref{thm:chunk}}
Summing $g(c_0+ki)$ over $i\le m$ gives $m[g_0+\lambda c_0]+\tfrac12\lambda km(m+1)$, and summing $g(c_0+s+ki)$ gives the same plus $\lambda sm$. Substituting into \eqref{eq:gd} with $n=L/m$ agents gives \eqref{eq:gdm}. Dropping terms constant in $m$, the per-step cost is $f(m)=\alpha m/2+h/m$ with $\alpha=\lambda(k-2s/L)$, so $f'(m)=\alpha/2-h/m^{2}$ and $f''(m)=2h/m^{3}$. For $h>0$ and $\alpha>0$, $f$ is strictly convex with minimiser \eqref{eq:mstar}, clipped to $[1,L]$, and a convex function restricted to the divisors of $L$ is minimised at one of the two divisors that bracket its continuous minimiser. For $h>0$ and $\alpha\le0$, $f$ is decreasing and the minimum is at $L$. For $h=0$, $f$ is linear and is minimised at an endpoint. Integers that do not divide $L$ change the objective through the final chunk and are compared through \eqref{eq:gd}. With $o>0$, \eqref{eq:deltao} gives $\Delta_o\approx(o/m)\sum_{i\le m}|\partial_b g(c_0+s+ki,b)|$ to first order, which is $\beta o$ when the sensitivity is roughly constant at $\beta$ across the chunk's loads, a constant per handoff that adds to $h$ in the first-order condition. \qed

\subsection*{Proposition~\ref{prop:summary}}
The objective $g_0+\lambda(c_0+s+k)+h(s)$ is strictly convex in $s$ as a sum of a linear and a strictly convex function, so an interior stationary point $-h'(s)=\lambda$, which exists by hypothesis, is its unique minimiser. \qed

\subsection*{Corollary~\ref{cor:verified}}
The first call uses $b$ and each of the $L-1$ boundaries reserves $o+b_g+o_v+b_r=b$, so the total is $bL$ even if every retry is used. Splitting a generator failure into undetected and detected cases gives failure probability $T(e,q_{\mathrm{eff}},e_r)$ for each fresh step, hence log-loss $\ell_v$, and with the additive handoff tax this gives $G_{\mathrm{D},v}$. Its advantage over the single agent is $D_v(L)=\sum_{t=2}^{L}[g(c_0+kt,b)-\ell_v-h]$ with $D_v(1)=0$, whose increments are nondecreasing, so the proof of Theorem~\ref{thm:crossover} applies verbatim. The unverified system with the same summary reservation has fresh-step loss $g(c_f,b-o)$, so unverified minus verified log-loss is $(L-1)[g(c_f,b-o)-\ell_v]$, which is nonnegative at every depth under the stated condition. \qed

\subsection*{Theorem~\ref{thm:vote}}
The indicators are conditionally i.i.d.\ $\mathrm{Bernoulli}(1-p)$ given $p\sim\mu$ by assumption, which gives \eqref{eq:vote}. For fixed $p<\nicefrac12$ the binomial tail tends to $0$, for $p>\nicefrac12$ to $1$, and for $p=\nicefrac12$ to $\nicefrac12$, and dominated convergence gives the limit. Under the common-shock law $\mu=\rho\eps\,\delta_1+\rho(1-\eps)\,\delta_0+(1-\rho)\,\delta_\eps$, so the mass above one half is $\rho\eps$, and \eqref{eq:voteshock} follows by conditioning on the shock. Monotonicity in $N$ for the $\delta_\eps$ component holds for $\eps<\nicefrac12$ and the other components do not depend on $N$. Exponential decay at $\rho=0$ is a Chernoff bound. The selector case replaces the majority event by ``at least one correct'', which for independent samples has probability $1-\eps^{N}$. \qed

\subsection*{Corollary~\ref{cor:cantelli}}
For conditionally i.i.d.\ indicators, $\mathrm{Cov}(X_i,X_j)=\mathrm{Var}(p)$, so $\sigma^{2}=\rho\,\eps(1-\eps)$. Cantelli's inequality gives $\mu(p-\eps\ge t)\le\sigma^{2}/(\sigma^{2}+t^{2})$ for $t>0$, and with $t=\nicefrac12-\eps$ the left side is $\mu(p\ge\nicefrac12)\ge F$. Equality in Cantelli holds for the two-point law with atoms at $\eps+t=\nicefrac12$ and $\eps-\sigma^{2}/t$, which is a valid law on $[0,1]$ iff $\eps-\sigma^{2}/t\ge 0$, i.e.\ $\rho\le(\nicefrac12-\eps)/(1-\eps)$, and moving the upper atom just above $\nicefrac12$ makes $F$ approach the bound. For the common-shock law, $\rho\eps\le F_{\max}$ with equality only at $\rho=0$, since for $\rho>0$ the inequality reduces to $\rho\eps\,t^{2}<\sigma^{2}(1-\rho\eps)$, i.e.\ $t^{2}<(1-\eps)(1-\rho\eps)$, which holds because $t^{2}<\nicefrac14<(1-\eps)^{2}\le(1-\eps)(1-\rho\eps)$ for $\eps<\nicefrac12$. \qed

\subsection*{Theorem~\ref{thm:parseq}}
By Theorem~\ref{thm:vote}, $\Evote(N)\to F$. By the saturation hypothesis, $\eps(c,Nb)\to\eps_\infty(c)$. If $F<\eps_\infty(c)$ the two limits are separated, so for all large $N$ the vote failure is below the sequential failure, and symmetrically in the other case. \qed

\section{Numerical verification and materials}
\label{app:numerics}

The closed forms were checked against brute-force evaluation of \eqref{eq:gsa} and \eqref{eq:gd}. Over 200 random draws of $(g_0,\lambda,k,s,h,c_0)$ with $\lambda\in[10^{-5},5\times10^{-4}]$, $k\in[50,2000]$, $s\in[0,3k]$, $h\in[0,0.5]$, the smallest depth at which $D(L)>0$ equals \eqref{eq:lstar} on every draw. Corollary~\ref{cor:flat} was checked exhaustively on all 2,358 combinations of $L\le 200$, $m\in\{1,2,5,10\}$ with $m\le L$, and $h\in\{0,0.01,0.1\}$ at a flat $g=0.02$, remainder chunks included, with no violation. The chunk-size formula agrees with a grid minimiser of the continuous per-step cost to a relative error of $1.7\times10^{-5}$ on the 91 of 100 draws with $m^{*}\ge 1$. Theorem~\ref{thm:chunks} was checked on the saturating curve $g(c)=0.2-0.18e^{-c/10}$ with $c_0=s=0$, $k=1$, and $h=0.204$, where $a_\infty=0.163<h<A_\infty=1.71$: $m=1$ never wins within 2,000 steps, and $m=2,3,5$ first win at depths 26, 15, and 9 and win at every depth from 28, 17, and 12, since a shorter final chunk can briefly reverse the sign. The identity in its proof matches direct evaluation, including remainder chunks, to within $10^{-13}$ over 200 random saturating curves. The threshold in Theorem~\ref{thm:parseq} was checked for $\eps(b)=0.08+0.22e^{-b}$ at $b=1$ under the common-shock law: the vote beats the sequential agent at every $N\in\{3,5,11,51,201\}$ for $\rho\in\{0,0.2\}$ and loses at every $N$ for $\rho\in\{0.5,0.8\}$, consistent with $\rho^{*}=0.497$.

Code and data are at \url{https://github.com/akshitanchan/attention-handoff-tax}. The repository contains the verification script, the calibration script (problem generator, filler generator, prompt, grader, resumable runner), the run configuration, the stored responses for all 1,000 pilot calls (final 200 characters each), the per-load summaries, the re-grading script that separates arithmetic from format failures, the script that fits the pilot curve and draws Figure~\ref{fig:calibration}, and, for Section~\ref{sec:exp}, the experiment script (ledger generator, both runners, the handoff-tax runner, the analysis), the per-step records for all 40,600 calls with finish reasons and token counts, the run configurations, the initial analysis output, and the exposure-matched analysis of Section~\ref{sec:exp} with its bootstrap and sensitivity runs. For the pilot, the serving stack was Ollama 0.30.8 with the packaged \texttt{gpt-oss:20b} model (Ollama ID \texttt{17052f91a42e}, 20.9B parameters, MXFP4) behind a derived model with a 71,680-token window (ID \texttt{eec2acdf0143}). The \texttt{ollama show} and \texttt{ollama list} output, captured when the files were assembled, is included with the run configuration.

\paragraph{Use of AI tools.}
The authors directed the research and conducted all experiments. AI tools (Anthropic's Claude and OpenAI's ChatGPT) were used as supplementary tools for checking and writing assistance. All results and claims were independently verified by the authors, who take full responsibility for the work.

\bibliographystyle{unsrtnat}
\bibliography{references}

\end{document}